%% file: Main.tex
\documentclass[pdflatex,sn-mathphys-num]{sn-jnl}

 \usepackage{algorithmicx} 
 \usepackage{placeins}
 \usepackage{xurl}
 \usepackage{algpseudocode} 
\usepackage{amsfonts} 
\usepackage{rsfso} 

\usepackage{svg}
\usepackage{xfrac}
\usepackage{graphicx}
\usepackage{epstopdf}
\usepackage{amsmath}
\usepackage{stfloats}
\usepackage{graphicx}%
\usepackage{multirow}%
\usepackage{amsmath,amssymb,amsfonts}%
\usepackage{amsthm}%
\usepackage{mathrsfs}%
\usepackage[title]{appendix}%
\usepackage{xcolor}%
\usepackage[T1]{fontenc}
\usepackage{lmodern}
\usepackage{mathrsfs}

\usepackage{textcomp}%
\usepackage{manyfoot}%
\usepackage{booktabs}%
\usepackage{algorithm}%
\usepackage{algorithmicx}%
\usepackage{algpseudocode}%
\usepackage{listings}%
\usepackage[printonlyused,withpage,nolist]{acronym}

\theoremstyle{thmstyleone}%
\theoremstyle{thmstyletwo}%

\theoremstyle{thmstylethree}%

\begin{document}

\title[Article Title]{Earthquake body-wave extraction using sparsity-promoting polarization filtering in the time–frequency domain}








\author*[1,4]{\fnm{Hamzeh} \sur{Mohammadigheymasi}}\email{hmohammadigheymasi@fas.harvard.edu}
\equalcont{These authors contributed equally to this work.}
\author[2]{\fnm{Bahare} \sur{Imanibadrbani}}
\equalcont{These authors contributed equally to this work.}
\author[3]{\fnm{Ali} \sur{Gholami}}
\equalcont{These authors contributed equally to this work.}
\author[2]{\fnm{Ahmad} \sur{Sadidkhouy}}
\equalcont{These authors contributed equally to this work.}

\affil*[1]{\orgdiv{Department of Earth and Planetary Sciences}, \orgname{Harvard University}, \orgaddress{\street{20 Oxford Street}, \city{Cambridge}, \postcode{02138}, \state{Massachusetts}, \country{USA}}}

\affil[2]{\orgdiv{Institute of Geophysics}, \orgname{University of Tehran}, \orgaddress{\street{End of North Kargar St. (Amirabad Ave.)}, \city{Tehran}, \postcode{1435944411}, \state{Tehran}, \country{Iran}}}

\affil[3]{\orgdiv{Institute of Geophysics}, \orgname{Polish Academy of Sciences}, \orgaddress{\street{Księcia Janusza 64}, \city{Warsaw}, \postcode{01-452}, \state{Masovian}, \country{Poland}}}

\affil[4]{\orgdiv{Atmosphere and Ocean Research Institute}, \orgname{The University of Tokyo}, \orgaddress{\street{5-1-5 Kashiwanoha}, \city{Kashiwa}, \postcode{277-8564}, \state{Chiba}, \country{Japan}}}



\abstract{Seismic waves generated by earthquakes consist of multiple phases that carry critical information about Earth’s internal structure as they propagate through heterogeneous media. Each seismic phase follows its own propagation path and sampling depth, bringing constraints from different regions of the Earth such as the crust, mantle, or even the outer and inner cores. The choice of phase for analysis therefore depends on the study target and the scientific objective: surface waves are particularly suited for imaging shallow, large-scale structures, whereas body waves provide higher-resolution information at depth, inaccessible to surface-wave methods. However, a persistent challenge in body-wave studies is that the relatively low-amplitude P and S arrivals are often obscured by stronger, slowly attenuating surface waves that overlap in both time and frequency. Although body waves typically contain higher frequency content, their spectral overlap with surface waves limits the effectiveness of conventional filtering approaches. Addressing this issue requires advanced signal processing techniques. One such method, Sparsity-Promoting Time–Frequency Filtering (SP-TFF; Mohammadigheymasi et al., 2022), exploits high-resolution polarization characteristics in the time–frequency domain to separate seismic phases. SP-TFF combines amplitude, directivity, and rectilinearity constraints to enhance phase discrimination. In this study, we further develop SP-TFF by designing a filter set specifically tailored to isolate body-wave arrivals that are otherwise masked by high-amplitude surface waves. The directivity filters are constructed based on the predicted incidence of seismic rays from the earthquake hypocenter to each station, enabling focused extraction of the incoming body wave energy and suppression of interfering phases, including surface waves and scattered wavefields. We demonstrate the method using both synthetic tests and waveform data from the $M_w 7.0$ Guerrero, Mexico, earthquake of September 8, 2021 (depth 21.8 km, reverse-thrust faulting), recorded by stations of the United States National Seismic Network (USNSN). Our results show that SP-TFF provides a robust computational framework for automated body-wave extraction, integrating polarization-informed filtering into seismological data processing pipelines. The approach is scalable to large waveform datasets and can enhance both real-time and retrospective seismological analyses, positioning it as a valuable informatics tool for Earth science. The codes required to reproduce the synthetic and observational examples are openly available on GitHub for the broader geoscience community.}




\maketitle

\section{Introduction}

Seismic waves are elastic disturbances generated by earthquakes, volcanic activity, or artificial sources, propagating through the interior of the Earth as vibrations of mechanical particles \citep{aki2002quantitative}. The propagation of the seismic wavefield is governed by the elastodynamic wave equation, with the original source function modified by interactions with the Earth's heterogeneous three-dimensional structure. These interactions generate different seismic phases and impart additional characteristics such as transmission, reflection, diffraction, dispersion, and attenuation. Ultimately, the full wavefield carries critical information about the elastic and structural properties of the subsurface, which can be retrieved once it is recorded on the Earth's surface by seismometers, highly sensitive instruments designed to measure ground motion \citep{shearer2019introduction}.

In practice, this physical foundation has motivated the development of a wide range of seismological methods that often focus exclusively on specific seismic phases, relying on first-order estimates of wave propagation patterns and arrival times \citep{kennett2001seismic}. This has further underscored the importance of developing robust techniques for extracting and separating individual seismic phases from the fully recorded wavefield, including applications to the separation of the upward and downward wavefield \citep{heravi2012curvelet}, shear-waves extraction passed from the lithosphere \citep{mohammadigheymasi2021sparsity}, the extraction of surface wave phases from ambient seismic noise for time-lapse monitoring \citep{nimiya2017spatial}, and body and coda wave discrimination \citep{mohammadigheymasi2022sparsity, schimmel2011polarized},
In particular, the extraction of high-frequency body waves has become an important objective in seismology due to their strong sensitivity to fine-scale velocity perturbations in relatively deeper structures \citep{brenguier2019train}. Extracting such phases, including direct, refracted, and reflected body waves, from seismograms is challenging, as their signals are typically masked by the dominant energy of surface waves, which decay more slowly with distance and therefore dominate body wave amplitudes \citep{draganov2013seismic}. Despite these difficulties, advances in phase extraction techniques have demonstrated that body wave phases can be separated from the continuous seismic wavefield \citep{ventosa2015extraction}. 

\cite{tonegawa2013extraction} proposed a processing scheme to extract Moho-reflected PpPp phases emerging in the teleseismic P coda using a deconvolution approach, in which the source wavelet of teleseismic P waves was estimated from vertical component records of a seismic array through non-linear waveform analysis. Building on passive approaches, \cite{nakata2014body} applied seismic interferometry to ground motions from clusters of regional earthquakes, retrieving direct and reflected plane waves that provide structural information. Subsequently, \cite{nakata2015body} demonstrated the retrieval of diving P and refracted waves from ambient noise records of a dense urban network in Long Beach, California. Similarly, \cite{ventosa2015extraction} tackled the extraction of PcP—a weak phase often hidden in the P coda—by employing a data-independent strategy based on the slant-stacklet transform and applying it to USArray data. Extending interferometric concepts, \cite{taylor2016crustal} constructed stacked autocorrelograms of ambient seismic noise to image body wave reflections, while \cite{xia2016extraction} reported clear observations of triplicated PKP branches (df, bc, ab) from stacked empirical Green’s functions obtained by correlating noise records across dense seismic arrays in South America and China. Brenguier et al. (2016) and Nakata et al. (2016) further demonstrated the temporal stability of direct virtual body waves between arrays at Piton de la Fournaise volcano, paving the way for continuous, passive monitoring of ballistic phases \citep{brenguier2020noise}. More recently, \cite{saygin2017retrieval} retrieved the P-wave reflectivity response of a thick sedimentary basin beneath Jakarta, Indonesia, from phase-weighted stacks of ambient-noise autocorrelations recorded on vertical components.



\citep{schneider2018improvement} developed and applied a frequency–wavenumber (f–k) filter to enhance data quality and detect small phases, such as PP precursors, in vespagrams. \cite{brenguier2019train} demonstrated that vehicle traffic, one of the most energetic and permanent anthropogenic seismic sources, can be exploited to reconstruct high-frequency body waves propagating across the San Jacinto Fault in Southern California to depths of several kilometers. Extending this concept, \cite{dales2020virtual} identified highway and railroad traffic as primary sources of high-frequency body-wave energy and showed that selective stacking of cross-correlation functions during vehicle and train passages yields clear body-wave arrivals. Similarly, \cite{zhang2020body} applied ambient-noise cross-correlation to dense seismic arrays in northeast China and successfully retrieved body-wave signals. Working with data from a dense array in the Dehdasht region of southwestern Iran, \cite{riahi2021simultaneous} simultaneously recovered high-frequency body waves and surface waves from the correlated noise field. To address synthetic passive signals, \cite{hariri2021retrieving} proposed and tested a Radon correlation approach for body-wave extraction. In parallel, \cite{johnson2021application} evaluated the performance of a publicly accessible CNN-based P-phase picker across multiple seismic networks of varying scale and instrumentation deployed to monitor long wall coal mining activity.

Improving the resolution of body-wave phase extraction remains an important and active area of research in seismology, requiring the development of increasingly sophisticated methods. In this study, we implement a high-resolution time–frequency filter for extracting earthquake body-wave phases using the Sparsity-Promoting Time–Frequency Filtering (SP-TFF) technique recently introduced by \cite{mohammadigheymasi2022sparsity}. This approach builds upon the Sparsity-Promoting Time–Frequency Representation (SP-TFR) framework, providing a refined decomposition of seismic wavefields in the time–frequency domain based on particle-motion characteristics, thereby enabling the separation of distinct seismic phases. This decomposition further allows the application of rectilinearity, directivity, and amplitude attributes in the time–frequency domain for either extracting or suppressing different seismic phases. Focusing on the propagation pattern and ray path of body waves, we employ ray tracing to estimate the incidence angle of the desired phases and to design directivity filters in the time–frequency domain. These filters capture the energy along the target direction and maximize the extraction of the wavefield associated with those phases. 

Our methodology builds on ideas first introduced at the EGU General Assembly \cite{imanibadrbani2022body}.
We briefly review the methodology and provide supplementary analysis of the underlying assumptions. We then schematically illustrate the directivity filter design. Finally, we present an optimal iterative strategy for solving the SP-TFR, based on the Fast Iterative Shrinkage–Thresholding Algorithm (FISTA). 
We organize the article as follows: first, we review the \ac{SP-TFF} methodology in Sec. \eqref{methodo}, then, we discuss setting up filer parameters for body waves extraction in Sec. \eqref{sec_body}. Afterward, in Sec. \eqref{sec_body} numerical examples with high-quality synthetic data and real data are presented to show the proposed method's performance on body wave extraction. Finally, we provide a discussion on the results and conclude the paper.

\subsection{Polarization analysis in the TF-domain} \label{methodo}

Noteworthy, we use a continuous notation in this section. The \ac{TF}-domain polarization parameters of particle motion, recorded by three-component seismic data, are defined for a continuous three-component signal
\begin{equation} \label{time_cont}
    \mathbf{X}(t) = 
    \begin{bmatrix}
        x_1(t) \\
        x_2(t) \\
        x_3(t)
    \end{bmatrix}, 
    \qquad t \in \mathbb{R},
\end{equation}
where $x_i(t)$ denotes the $i$th component of ground motion.

The \ac{TF} coefficients of the three components are obtained via a continuous TF operator $\mathcal{TF}\{\cdot\}$:
\begin{equation} \label{TF_cont}
    \mathbf{X}(t,f) = 
    \mathcal{TF}\!\left\{\mathbf{X}(t)\right\} 
    = 
    \begin{bmatrix}
        X_1(t,f) \\
        X_2(t,f) \\
        X_3(t,f)
    \end{bmatrix}, 
    \qquad (t,f) \in \mathbb{R}\times\mathbb{R},
\end{equation}
where $X_i(t,f)$ is the TF representation of $x_i(t)$. Several well-established methods exist for computing \eqref{TF_cont}, including the continuous wavelet transform \citep{mallat1999wavelet}, the Wigner–Ville distribution \citep{flandrin1998time}, the Short-Time Fourier Transform (\ac{STFT}) \citep{1162950}, and the Stockwell Transform (\ac{ST}) \citep{stockwell2007basis}.

Taking the \ac{STFT} as an example, the continuous TF representation of $x_i(t)$ is given by
\begin{equation} \label{STFT_cont}
    X_i^{\text{STFT}}(t,f)
    = \int_{-\infty}^{\infty} 
        x_i(\tau)\, w(\tau - t)\, 
        e^{-\,\jmath\,2\pi f \tau}\, 
    d\tau,
\end{equation}
where $w(\cdot)$ is an analysis window, $f$ is frequency (Hz), and $\jmath^2 = -1$.

Accordingly, the three-component TF representation can be expressed as
\begin{equation}
    \mathbf{X}^{\text{STFT}}(t,f) =
    \begin{bmatrix}
        X_1^{\text{STFT}}(t,f) \\
        X_2^{\text{STFT}}(t,f) \\
        X_3^{\text{STFT}}(t,f)
    \end{bmatrix},
\end{equation}

where
\begin{equation} \label{Gauss_cont}
    w(\tau - t) = \frac{1}{\sigma \sqrt{2\pi}} 
    \exp\!\left[-\frac{(\tau - t)^2}{2\sigma^2}\right]
\end{equation}
represents a Gaussian window with standard deviation $\sigma$, centered at time $t$. 
In this continuous formulation, $t \in \mathbb{R}$ denotes time, and $f \in \mathbb{R}$ frequency.

Returning to our main topic, by integrating the \ac{TF} coefficients of the three components specified in \eqref{TF_cont}, 
we can calculate the TF-dependent polarization characteristics of 3D motion. 
This is done in terms of the eigenvectors 
$\mathbf{u}_1(t,f)$, $\mathbf{u}_2(t,f)$, $\mathbf{u}_3(t,f)$ 
and eigenvalues 
$\lambda_1(t,f)$, $\lambda_2(t,f)$, $\lambda_3(t,f)$, 
which are obtained by solving
\begin{equation} \label{eig_tff_cont}
    \big(\mathbf{C}(t,f) - \lambda_i(t,f)\,\mathbf{I}\big)\,\mathbf{u}_i(t,f) = 0.
\end{equation}

The cross-spectral (covariance) matrix $\mathbf{C}(t,f)$ is defined at each time-frequency location:
\begin{equation} \label{acc_cont}
    \mathbf{C}(t,f) = \big[C_{ij}(t,f)\big]_{i,j=1}^3 \in \mathbb{C}^{3\times 3}.
\end{equation}

Here, $i,j=1,2,3$ index the different component pairs of the signal. 
The individual TF-domain cross-spectral terms are computed as
\begin{equation} \label{single_f_abs_cont}
    C_{ij}(t,f) = X_i(t,f)\,X_j^*(t,f),
\end{equation}
where $^*$ denotes the complex conjugate. 
In practice, suitable normalization factors are applied depending on the chosen TF decomposition 
(e.g., $\text{STFT}$, wavelet, or Stockwell transform).  

\begin{figure}[h!]
		\centering
		\includegraphics[width=0.9\textwidth]{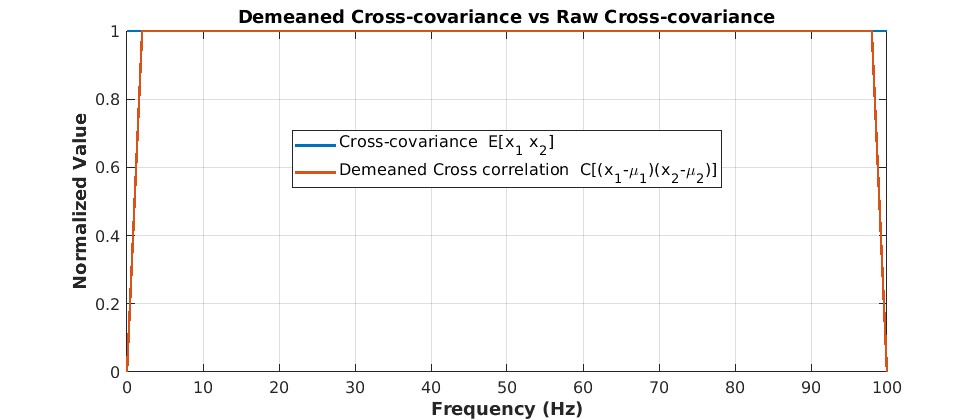} 
		\caption{This figure shows the value of bias introduced by  assuming a weakly stationary condition for all the components of the time series, implying $ \mu_i\cong 0, \hspace{0.1cm} i=1,2,3$. This assumption essentially reduces the definition of elements in the covariance matrix \eqref{acc_cont} to auto- and cross-correlations. We have conducted a simulation to show the impact this approximatiion on the accuracy of estimations for different frequencies. The blue curve shows the exact value of cross-covariance between While the assumption simplifies calculations, the resulting matrix $\boldsymbol{C}$ might not entirely reflect the variability and dynamics inherent in the original time series, particularly for low- and high-frequency extremes. Here, we conduct a numerical simulation to assess the effect of this approximation on a range of frequencies from 0 to the Nyquist frequency.   
        }
		\label{fig:appr}
	\end{figure}

In the computation of \eqref{eig_tff_cont} using \eqref{acc_cont}, 
a weakly stationary assumption is often made for the three components of the time series, 
implying $\mu_i \approx 0$ for $i=1,2,3$. 
This assumption essentially reduces the definition of the covariance matrix elements to auto- and cross-correlations. 
While this simplifies calculations, the resulting $\mathbf{C}(t,f)$ may not fully capture the variability 
and dynamics of the original signal, particularly at very low and very high frequencies. 
To illustrate the effect of this approximation, we perform a numerical simulation of a monochromatic wave of frequency $f$, 
propagating in 3D space and recorded on two components of a sensor in the $X$ and $Y$ directions. 
The simulation covers a frequency range from 0 to the Nyquist frequency (100 Hz). 
The exact cross-covariance of the two components versus the detrended (demeaned) cross-correlation 
is shown in Fig.~\ref{fig:appr}.

As shown, the assumption provides a reasonable estimate of the true cross-covariance value, except in the very low-frequency range (<1 Hz) and at very high frequencies near the Nyquist frequency. In practice, seismic data naturally have a limited frequency band, and under this condition the assumption appears satisfactory in terms of accuracy.

\subsection{Enhancing the resolution by Sparsity-promoting TF-decomposition}

The \ac{STFT} representation \eqref{STFT_cont}, and similarly the \ac{ST} transform, 
can be expressed in matrix form as a linear system: 

\begin{equation} \label{linear}
\boldsymbol{x}= \boldsymbol{G}\boldsymbol{\alpha}, \hspace{1cm}
\boldsymbol{G}\in \mathbb{C}^{L\times L^2}, \hspace{1cm} 
\boldsymbol{\alpha} \in \mathbb{C}^{L^2\times 1},
\end{equation}

where $\boldsymbol{x}$ is the vector of observed time-domain samples, 
$\boldsymbol{G}$ is the forward operator, 
and $\boldsymbol{\alpha}$ is a vectorized form of the TF coefficients defined in \eqref{STFT_cont}. 
Further details about the construction of $\boldsymbol{G}$ can be found in \cite{vera2012microseismic, gholami2012sparse}. 
Since \eqref{linear} is an underdetermined system, infinitely many TF maps can represent the signal. 
The optimal map can be obtained through the inclusion of a priori information within a regularization framework \citep{vera2012microseismic, gholami2012sparse}.

A sparsity-promoting regularization selects a TF model with the fewest non-zero coefficients by solving

\begin{align} \label{unconstrained} 
	\boldsymbol{\alpha}=\arg \min_{\boldsymbol{\alpha}} 
	\frac{1}{2}\|  
	\boldsymbol{G}\boldsymbol{\alpha}-\boldsymbol{x}\|^2_2  
	+ \mu \| \boldsymbol{\alpha}\|_1,
\end{align}

where the $\ell_2$- and $\ell_1$-norms represent the fit-to-data and regularization terms, respectively, 
and $\mu>0$ controls the resolution of the TF map. 
When properly tuned, $\mu$ allows discrimination between closely spaced events in time and frequency while ensuring accurate reconstruction \cite{vera2012microseismic, gholami2012sparse}. 
Importantly, the use of the $\ell_1$ norm retains convexity of the cost function, enabling efficient optimization 
and avoiding local minima associated with non-convex penalties.

This problem is equivalently written in constrained form, following the Karush–Kuhn–Tucker (KKT) conditions \cite{gordon2012karush}:

\begin{align} \label{constrained}
\arg \min_{\boldsymbol{\alpha}}
\|\boldsymbol{\alpha}\|_1, 
\quad \text{subject to} \quad 
\frac{1}{2}\|\boldsymbol{G}\boldsymbol{\alpha}-\boldsymbol{x}\|^2_2 \leq \epsilon.
\end{align}

Here, $\epsilon$ controls the tolerance on the reconstruction error, preventing overfitting to Gaussian noise and improving robustness. Several algorithms can solve \eqref{unconstrained}, including Iteratively Reweighted Least Squares (IRLS) 
\cite{gheymasi2013local,gholami2016regularization}, 
Split-Bregman iterations \cite{gholami2012sparse,gheymasi2016robust}, 
and the Fast Iterative Shrinkage-Thresholding Algorithm (\ac{FISTA}) \citep{beck2009fast}. 
In this study, we adopt the \ac{FISTA} approach, which provided the most efficient solution. 
A pseudocode implementation is shown in Algorithm~\eqref{FISTA}.

\begin{algorithm}\label{FISTA}
\caption{STFT\_S\_IST}
\begin{algorithmic}[1]
\Function{STFT\_S\_IST}{x, y, z, s, tt, fff}
    \State N $\gets$ length(x)
    \State d1 $\gets$ x
    \State d2 $\gets$ y
    \State d3 $\gets$ z
    \State window $\gets$ 'Gaussian'
    \State window\_half\_length $\gets$ s/2
    \State G $\gets$ Gabor(window, window\_half\_length, length(x))
    \State $A \gets @(e1)$ G*e1
    \State $At \gets @(e1)$ G'*e1
    \State dh1 $\gets$ d1
    \State dh2 $\gets$ d2
    \State dh3 $\gets$ d3
    \State n\_it $\gets$ 400
    \State mu $\gets$ 2e-3
    \State lambda $\gets$ mu * max(abs(2*pi*At(d1))) * 1.2
    \State tfx $\gets$ zeros($N^2$,1)
    \State tfy $\gets$ zeros($N^2$,1)
    \State tfz $\gets$ zeros($N^2$,1)
    \For{j = 1 to n\_it}
        \State U1j $\gets$ tfx + mu * 2*pi * At(d1 - 2*pi * A(tfx))
        \State U2j $\gets$ tfy + mu * 2*pi * At(d2 - 2*pi * A(tfy))
        \State U3j $\gets$ tfz + mu * 2*pi * At(d3 - 2*pi * A(tfz))
        \State tfx $\gets$ soft\_complex(U1j, lambda)
        \State tfy $\gets$ soft\_complex(U2j, lambda)
        \State tfz $\gets$ soft\_complex(U3j, lambda)
        \State d1 $\gets$ d1 + (dh1 - real(2*pi*A(tfx)))
        \State d2 $\gets$ d2 + (dh2 - real(2*pi*A(tfy)))
        \State d3 $\gets$ d3 + (dh3 - real(2*pi*A(tfz)))
    \EndFor
    \State tfx $\gets$ reshape(tfx,N,N)
    \State tfy $\gets$ reshape(tfy,N,N)
    \State tfz $\gets$ reshape(tfz,N,N)
    \State recx $\gets$ real(2*pi*A(tfx))
    \State recy $\gets$ real(2*pi*A(tfy))
    \State recz $\gets$ real(2*pi*A(tfz))
    \State \Return tfx, tfy, tfz, recx, recy, recz
\EndFunction
\end{algorithmic}
\end{algorithm}

 \FloatBarrier

Subsequently, the derived sparsity-promoting time-frequency representation (SP-TFR) from \eqref{unconstrained} was applied to formulate an adaptive filter capable of extracting different seismic wave phases. A brief overview of the adaptive filtering procedure in the TF domain is presented in Subsection. \ref{pola_fil}.

Obtaining a high-resolution representation in the time-frequency domain has always been a big challenge. On this subject, combining SP-TFR with formulation of the EDPA responds favorably to this need of the scientific community. In the last decade, sparsity has been proved as a promising tool to solve inverse problems for a high-resolution solution which mathematically may be nonunique. It has successfully been applied to many areas of data processing and inversion, including deconvolution/deblurring, migration, tomography, interpolation, and Radon transform, just to name a few \cite{gholami2012sparse}. 
	Nevertheless, TFR is considered as an inverse problem in a way that the TF coefficients defined as a solution of a linear equation. Since the desired linear system is considered under-determined, there are infinitely many TF maps to represent the signal. An SP regularization enables selecting a TF model with minimum nonzero coefficients by solving a constrained optimization problem.

\subsection{Filter components and filter design} \label{pola_fil}

1) Rectilinearity Attribute: Rectilinearity is one of the three parameters used to distinguish between linear and elliptical particle motion. The degree of rectilinearity is
\begin{align} \label{rect}
Re(t,f)=\dfrac{\lambda _1(t,f)-\lambda _2(t,f)-\lambda _3(t,f)}{\lambda _1(t,f)}
\end{align}
defined as a rectilinearity measure to discriminate between the rectilinear motion of Love and body waves and the elliptical motion of Rayleigh waves. Hence, the rectilinearity filter is designed in the TF-domain as
\begin{align} \label{Psi_Re}
\Psi_{Re}(Re(t,f)) =
\begin{cases}
  1, & -1 < Re(t,f) < \alpha, \\[6pt]
  \cos\!\left(\dfrac{\pi\,(Re(t,f)-\alpha)}{2(\beta-\alpha)}\right), & \alpha < Re(t,f) < \beta, \\[10pt]
  0, & \beta < Re(t,f) < 1.
\end{cases}
\end{align}

2) Directivity Attribute: Another parameter to discriminate between different seismic phases is the directivity attribute. This parameter is based on the direction of particle motion. A directivity measure is defined as the absolute value of the dot product
of the first eigenvector with the base vectors
\begin{align} \label{directivity}
D_i(t,f) &= \left| \boldsymbol{u}_1^T(t,f) \cdot \boldsymbol{e}_i \right|, 
&& i \in \{T, R, Z\}.
\end{align}

Then normalizing the measure in the TF plane. Accordingly, the directivity filter is designed in the TF-domain as
\begin{align} \label{Psi_D}
\Psi_D\big(D_i(t,f)\big) &=
\begin{cases}
  1, & 0 < D_i(t,f) < \gamma, \\[6pt]
  \cos\!\left(\dfrac{\pi\,(D_i(t,f)-\gamma)}{2(\lambda-\gamma)}\right), & \gamma < D_i(t,f) < \lambda, \\[10pt]
  0, & \lambda < D_i(t,f) < 1,
\end{cases} \\[8pt]
& i \in \{T, R, Z\}. \nonumber
\end{align}

	\begin{figure}[h!]
		\centering
		\includegraphics[width=0.9\textwidth]{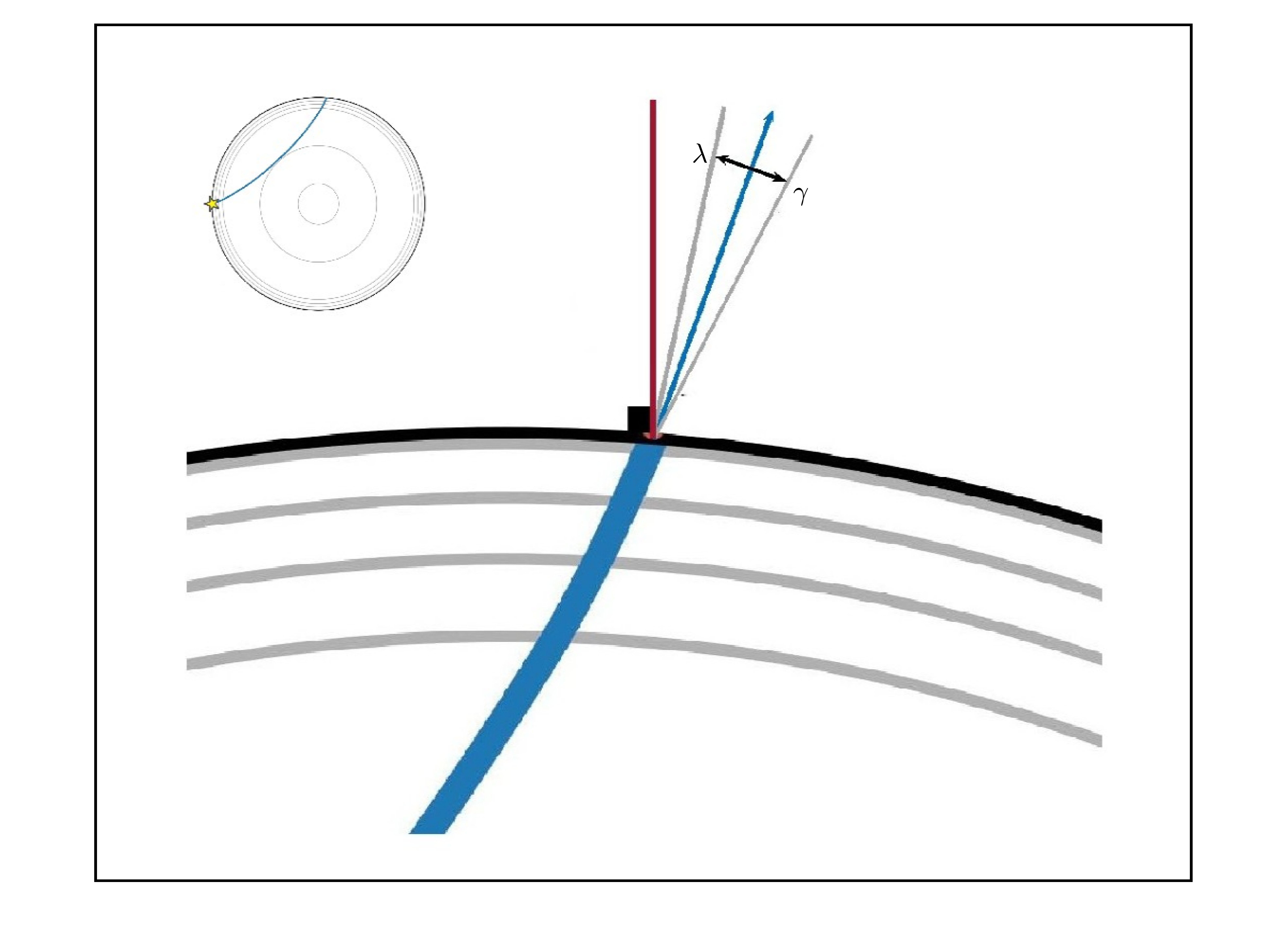} 
		\caption{This schematic figure shows the optimal design of the directivity filter design for a seismic wavefield propagating from the seismic source towards the seismic station. Ray tracing calculates the incident angle, and provides information for defining the filter pannel for passing the signal.        
        }
		\label{fig:Waveforms}
	\end{figure}

3) Amplitude Attribute: Given that the amplitudes of body waves are much smaller than those of surface waves, separating these waves from surface waves is a challenge worldwide. Thus, the amplitude parameter can be recognized as the most critical parameter to extract body waves from surface waves, specifically at low frequencies. The amplitude attribute is defined as
\begin{align} \label{amplitude}
A(t,f)=\sqrt{2\lambda_1(t,f)}
\end{align}
and the amplitude filter is designed in the TF-domain as
\begin{align} \label{Psi_A}
\Psi_A\big(A(t,f)\big) &=
\begin{cases}
  1, & 0 < A(t,f) < \zeta, \\[6pt]
  \cos\!\left(\dfrac{\pi\,(A(t,f)-\zeta)}{2(\eta-\zeta)}\right), & \zeta < A(t,f) < \eta, \\[10pt]
  0, & \eta < A(t,f) < 1.
\end{cases}
\end{align}

Integrating all these filters  the total time–frequency (TF) reject filter for suppressing a seismic phase is constructed by combining the rectilinearity, directivity, and amplitude filters as
\begin{equation} \label{total}
\boldsymbol{\Phi}_R(t,f) = 
1 - \Big\{1-\boldsymbol{\Phi}_{Re}(t,f)\Big\} 
\circ \Big\{1-\boldsymbol{\Phi}_{D}(t,f)\Big\} 
\circ \Big\{1-\boldsymbol{\Phi}_{A}(t,f)\Big\},
\end{equation}
where $\circ$ denotes the Hadamard (element-wise) product. 

Similarly, an extract filter for isolating a specific seismic phase can be defined as
\begin{equation} \label{extract}
\boldsymbol{\Phi}_E(t,f) = 1 - \boldsymbol{\Phi}_R(t,f).
\end{equation}

The filtering process is performed by element-wise multiplication of the chosen filter $\boldsymbol{\Phi}(t,f)$ 
with the SP-TFR of the three components. 
The filtered signal in the time domain is then reconstructed via the inverse transform formulation in \eqref{linear}, 
yielding the sparsity-promoting time–frequency filter.

\section{Results and Discussion}

\subsection*{Body wave extrcation from earthquake waveforms }\label{sec_body}

\begin{figure}[h!]
		\centering
		\includegraphics[width=0.9\textwidth]{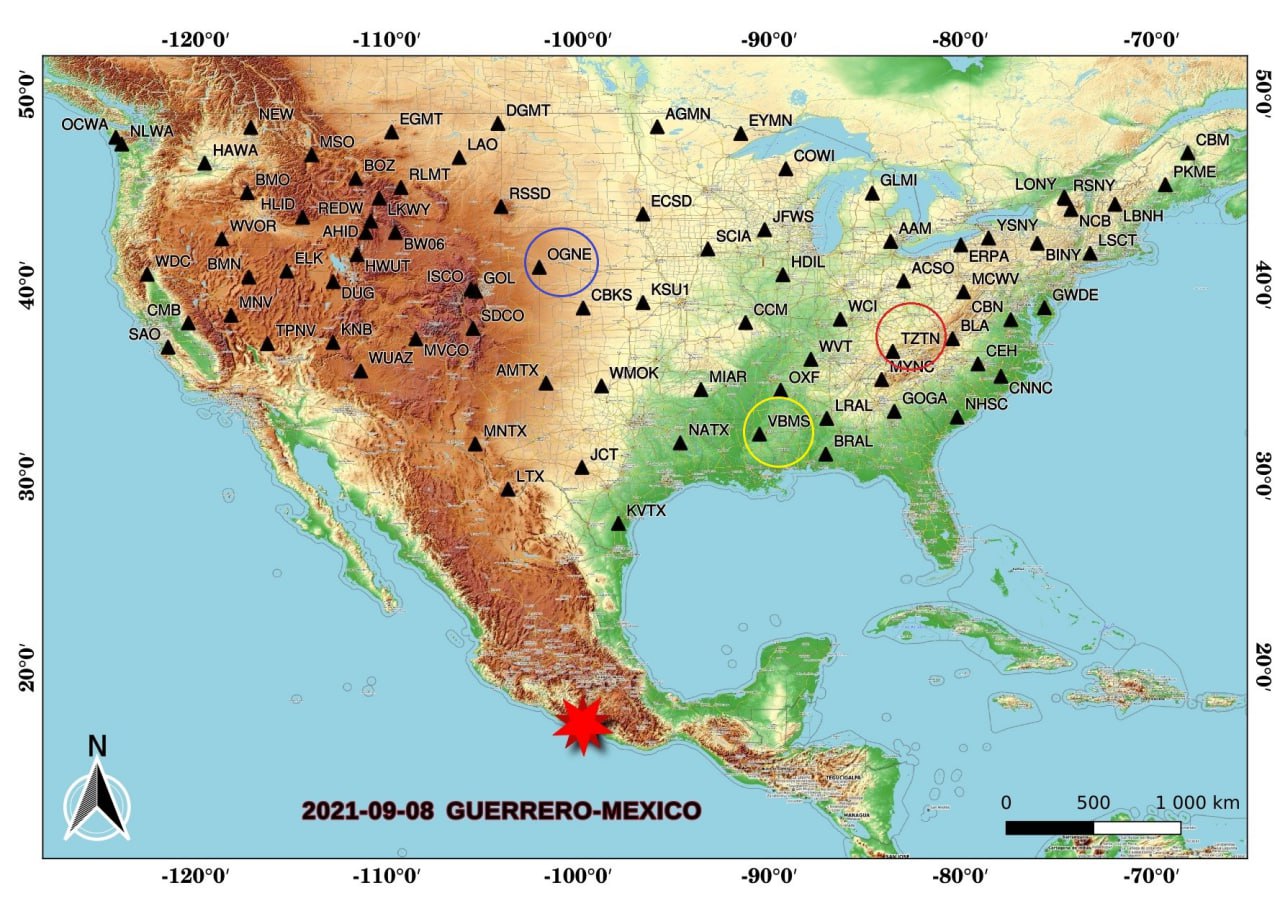} 
		\caption{The map shows the locations of the stations OGNE, VBMS, and TZYN among the USNSN, whose synthetic and real datasets recording the $M_w = 7.0$ earthquake that occurred at a depth of 21.8 km in Guerrero, Mexico, on 08-09-2021 have been processed in this study for body-wave extraction.       
        }
		\label{fig:network}
	\end{figure}

In this section, we demonstrate the performance of the proposed method in extracting body-wave phases from earthquake waveforms, using both synthetic and real data. For the synthetic tests, we model an earthquake of moment magnitude $M_w = 7.0$, occurred at a depth of 21.8 km in Guerrero, Mexico, on Wednesday, September 8, 2021, at 01:47:43 UTC, associated with a reverse-thrust fault. The analysis focuses on recordings from the TZNT and VBMS stations of the United States National Seismic Network (USNSN), evaluating the extraction of body waves from overlapping Rayleigh waves on the radial and vertical components (Fig. 		\eqref{fig:network}). For the real data case, we analyze recordings from the OGNE station of the same network.

	\subsection*{Synthetic examples}

    To generate synthetic data, we integrated the seismic network geometry and  hypocenteral parameters and focal mechanism information of the 2021 Guerrero earthquake. The 3-component synthetic waveforms of particle velocity  are enerated using using the AxiSEM library through the IRIS Synthetics Engine (Syngine) within the ObsPy software \cite{beyreuther2010obspy}. An spectral-element method generated wave propagation in a three-dimensional, non-elastic, anisotropic spherical model in the f135ak D-1 Earth model. We synthesize recordings for both  station VBMS to evaluate the effectiveness of the SP-TFF method in separating body waves from Rayleigh waves. The figure below shows the locations of the earthquake and station VBMS.

	The three-dimensional synthetic earthquake data were generated  The simulation was performed using the AxiSEM library through the IRIS Synthetics Engine (Syngine) within the ObsPy software. The generated data were processed as follows: first, detrending was applied; then, time decimation by a factor of 8 was performed to obtain a dataset with a sampling interval of 2 seconds. Next, the seismograms were rotated into the transverse--radial--vertical coordinate system. To make the simulation more realistic, Gaussian noise $n \in \mathbb{R}^{L \times 1}$ was added (bandpass filtered in the range 0.02--0.5 Hz) to achieve a signal-to-noise ratio (SNR) of 10. The waveforms recorded on the transverse, radial, and vertical components for the synthetic earthquake at station VBMS are shown in panels (a)--(c) of the figure below.
	\begin{figure}[h!]
		\centering
		\includegraphics[width=0.8\textwidth]{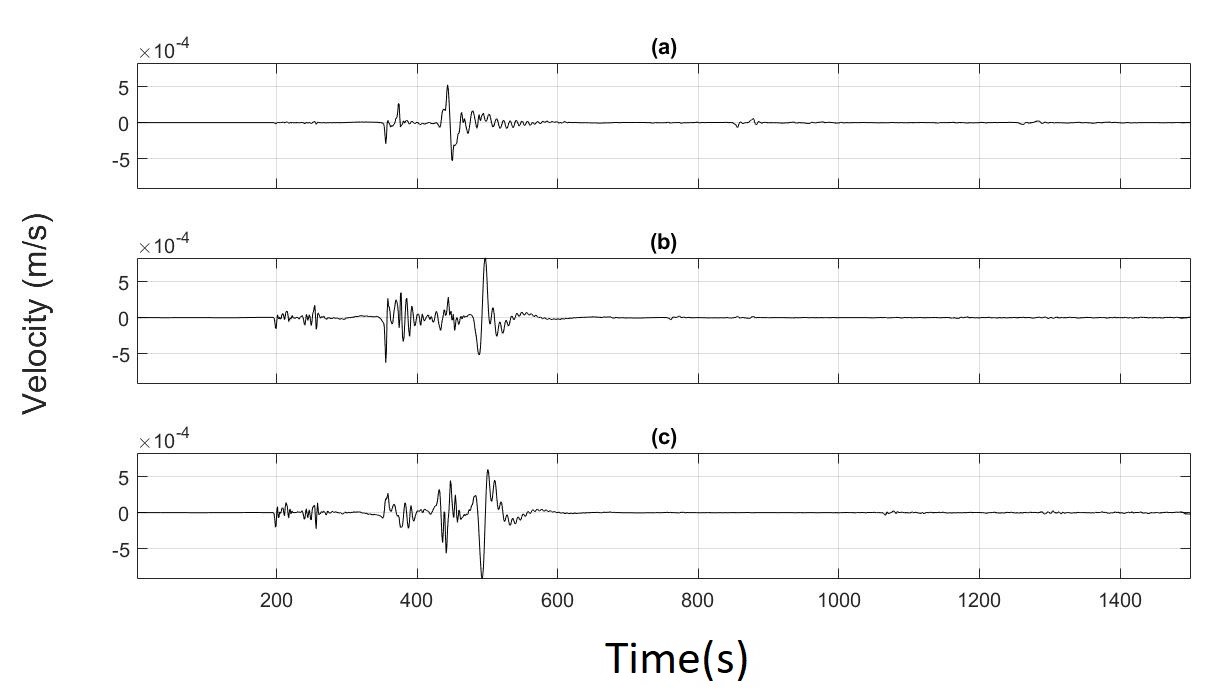} 
		\caption{Waveforms shown in panels (a)--(c) correspond to the transverse, radial, and vertical components of the three-dimensional synthetic earthquake dataset with a moment magnitude of $M_w = 7.0$ in Guerrero, Mexico, generated for station VBMS using the AxiSEM library.}
		\label{fig:Waveforms1}
	\end{figure}
	
	To ensure the efficiency of the SP-TFF method, its results are compared with those obtained using the Pingar method. In this approach, the Stockwell Transform (ST) is used to compute the time--frequency representation (TFR), and the time--frequency domain polarization parameters are obtained by fitting particle motion to a parametric ellipse using the combined TFR of the three components. More precisely, the set of polarization parameters, including $a(k,l)$ (length of the SM axis of the parametric ellipse), $b(k,l)$ (length of the Sm axis of the parametric ellipse), $I(k,l)$ (tilt angle of the ellipse relative to the horizontal plane at $(k,l)$), $\Omega(k,l)$ (azimuth of the major axis), and $\phi(k,l)$ (phase measured at the time of maximum displacement), are determined. Here, $l = 0, \dots, n$ and $k = 0, 1, \dots, L$ denote the time and non-negative frequency indices, respectively.
	
	The time--frequency representations of the radial, transverse, and vertical components of the synthetic earthquake data obtained by the ST method can be seen in panels (a), (c), and (e) of the figure below. The corresponding time--frequency representations for the SP-TFF method are shown in panels (b), (d), and (f).
	
	\begin{figure}[h!]
		\centering
		\includegraphics[width=1\textwidth]{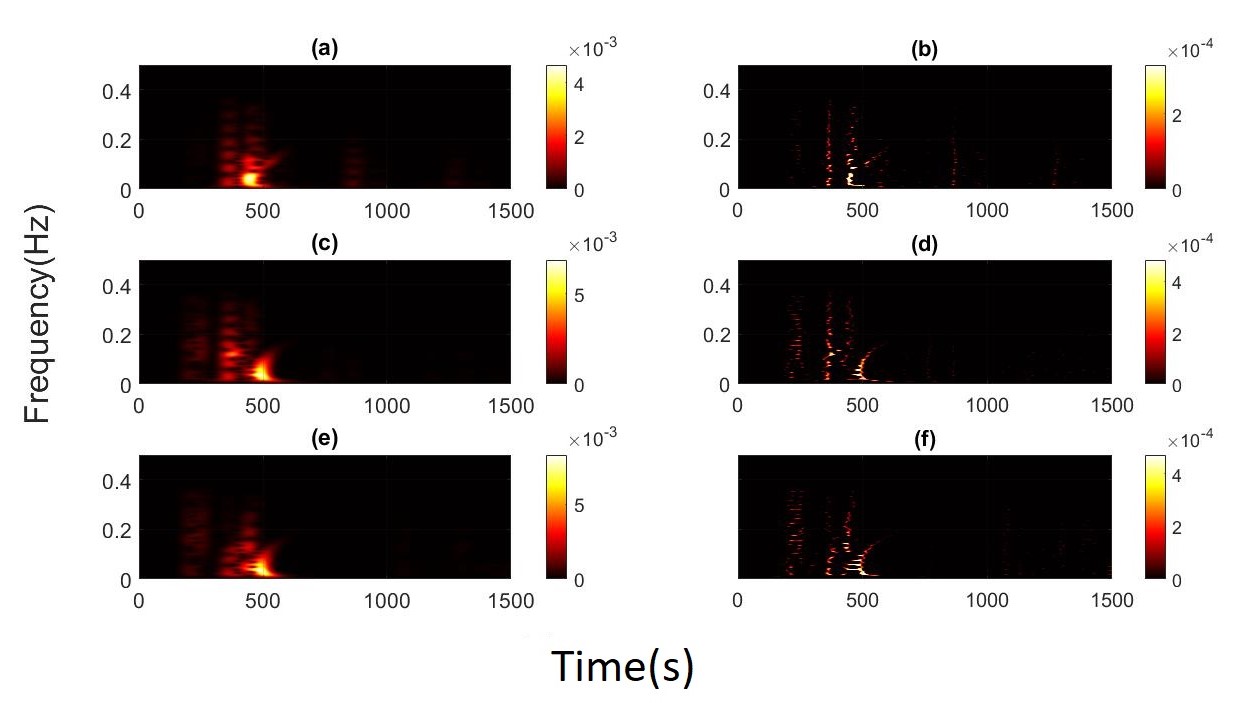} 
		\caption{Time--frequency representations of the synthetic dataset shown in Figure 2-4. Panels (a), (c), and (e) correspond to the radial, transverse, and vertical components obtained using the ST method, while panels (b), (d), and (f) show the corresponding components obtained using the SP-TFF method.}
		\label{fig:Waveforms11}
	\end{figure}

	It is clearly observable that the results obtained using the SP-TFF method are significantly more accurate than those obtained with the ST method proposed by Pingar (2006). It is also noted that the maximum amplitude in SP-TFF is larger than that in ST, while the energy dispersion in ST is considerably higher than in SP-TFF. As seen on the left side of the figure, with the ST method, waves are inseparably overlapping and cannot be distinguished. In contrast, the SP-TFF method (right side) separates seismic waves with high clarity and precision, allowing the distinct identification of seismic phases. The overall patterns of surface and body waves are discernible in both TFRs.
	
	The major and minor axes of the particle-motion ellipses are denoted as $SM$ and $Sm$, respectively. In panels (a) and (c), the major and minor axes obtained using Pingar’s method (2006) are shown, whereas panels (b) and (d) display the corresponding axes derived using Eigenvalue Decomposition Polarization Analysis (EDPA) applied to SP-TFF.  
	
	\[
	SM(k,l) = \sqrt{2 \lambda_1(k,l)}
	\]  
	\[
	Sm(k,l) = \sqrt{2 \lambda_2(k,l)}
	\]  
	
	As previously defined, $k$ and $l$ correspond to the frequency and time indices, respectively.
	\begin{figure}[h!]
		\centering
		\includegraphics[width=1\textwidth]{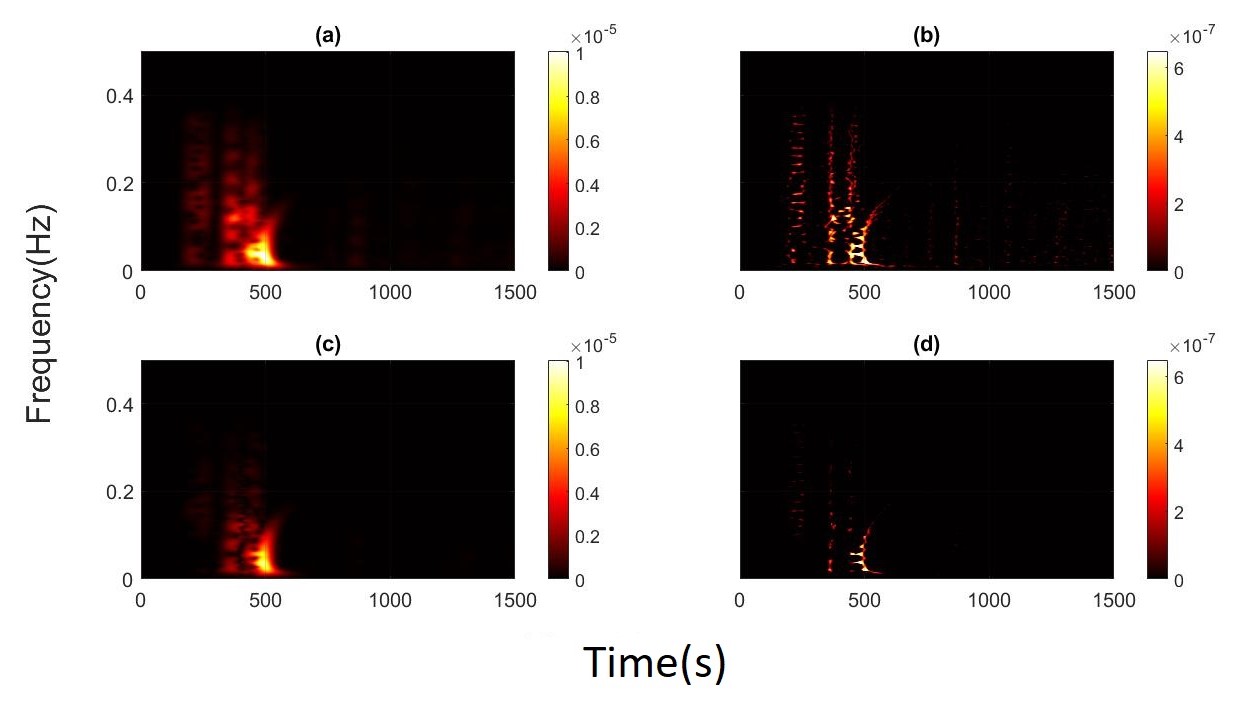} 
		\caption{Time--frequency representations of the major ($SM$) and minor ($Sm$) axes of particle motion for the synthetic dataset shown in Figure 2-4. Panels (a) and (c) correspond to the $SM$ and $Sm$ axes obtained using Pingar’s method (2006), while panels (b) and (d) show the $SM$ and $Sm$ axes derived from applying EDPA to the SP-TFF method.}
		\label{fig:Waveforms2}
	\end{figure}
	
	Next, by combining the time--frequency domain polarization parameters obtained from SP-TFF, the data are processed to filter out the Rayleigh wave phases. The polarization parameters in the time--frequency domain obtained from SP-TFF are used to define an adaptive filter for separating S and P waves as well as removing Love and Rayleigh waves.  
	
	To filter the Rayleigh wave phases, the directivity measure relative to the radial--vertical plane is calculated as:  
	
	\[
	D(k,l) = \sqrt{D_R(k,l)^2 + D_Z(k,l)^2}
	\]  
	
	Additionally, a rectilinearity filter is designed and applied to remove these waves. Accordingly, the SP-TFF components of the Rayleigh-wave-filtered data are shown in panels (b), (d), and (f) of Figure 5-4. As clearly observable, the Rayleigh waves on the radial and vertical components are removed with high accuracy, while the scattered energy of the body waves in the SP-TFF-filtered data remains on the radial and vertical components.
	\begin{figure}[h!]
		\centering
		\includegraphics[width=1.0\textwidth]{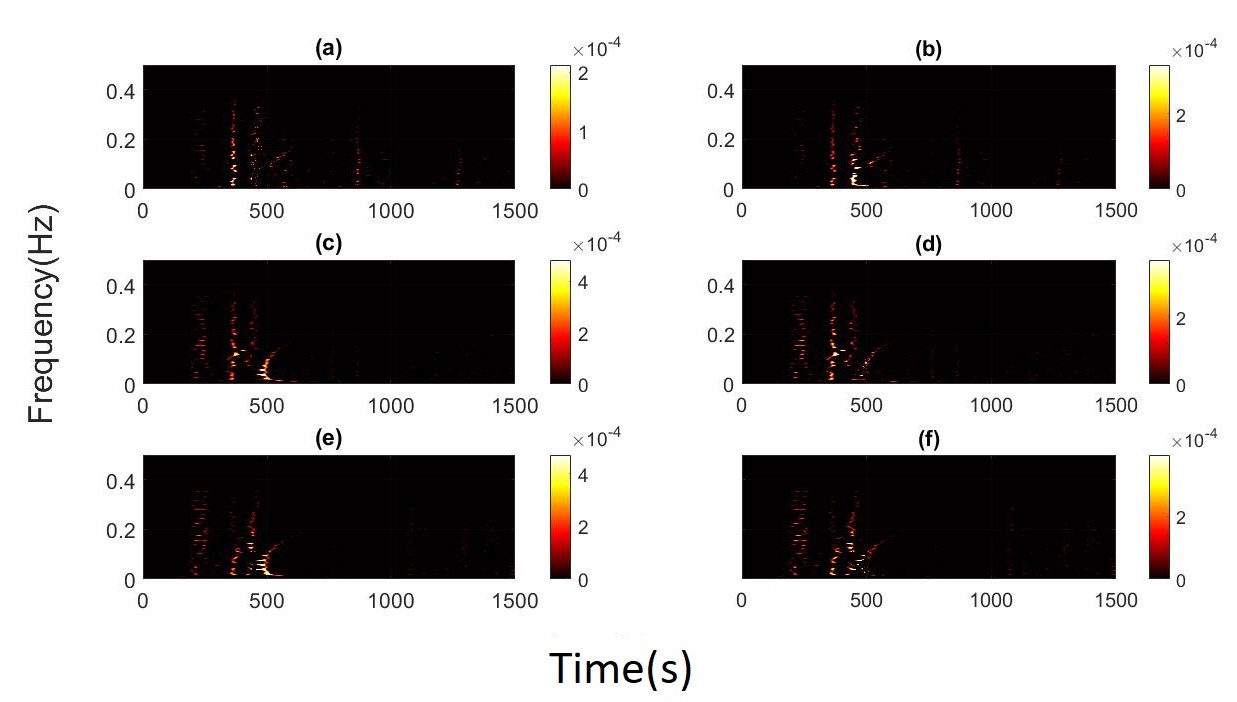} 
		\caption{Adaptive filtering for surface-wave removal and body-wave separation. Panels (a), (c), and (e) show the SP-TFR of Love-wave-filtered data on the transverse, radial, and vertical components, respectively. Panels (b), (d), and (f) display the SP-TFR of Rayleigh-wave-filtered data on the same components.}
		\label{fig:Waveforms22}
	\end{figure}
	
	The transverse, radial, and vertical components of the Rayleigh-wave-filtered data, reconstructed in the time domain, are shown in panels (b), (d), and (f) of the figure below, respectively.
	\begin{figure}[h!]
		\centering
		\includegraphics[width=1.0\textwidth]{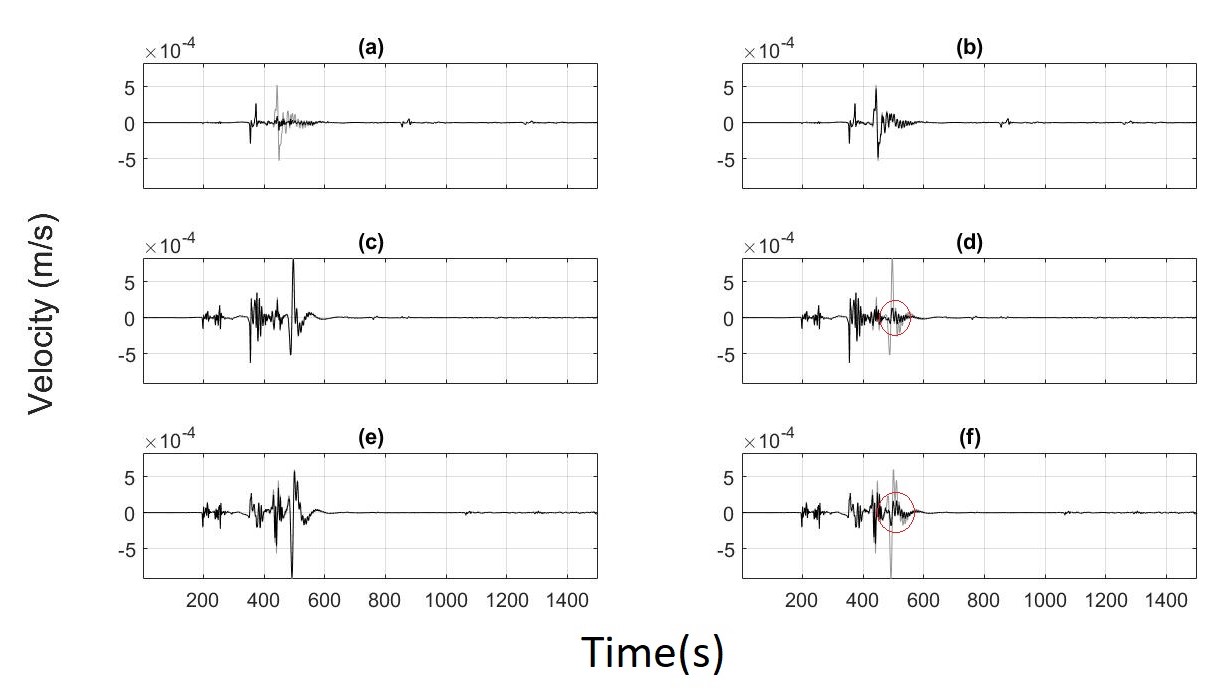} 
		\caption{Filtered surface waves and remaining body waves for the synthetic earthquake. The black waveforms in panels (a), (c), and (e) correspond to the transverse, radial, and vertical components of the Love-wave-filtered synthetic dataset shown in Figure 2-4, obtained using SP-TFF. Panels (b), (d), and (f) show the corresponding components for the Rayleigh-wave-filtered data. The gray waveforms in all panels represent the raw synthetic dataset.}
		\label{fig:Waveforms222}
	\end{figure}
	
	As shown, the SP-TFF method effectively filters the Rayleigh waves without affecting the body waves on the radial and vertical components, and has only a minor effect on other phases on the transverse component. As a final assessment, by applying Equation (3-33) to the SP-TFR, the Rayleigh-wave phases are extracted on all three components. Panels (d) and (f) of Figure 7-4 show the radial and vertical seismograms corresponding to the separated Rayleigh waves. It is noteworthy that the Rayleigh phases have been extracted from the complete seismograms with high accuracy.
	
	\begin{figure}[h!]
		\centering
		\includegraphics[width=1.0\textwidth]{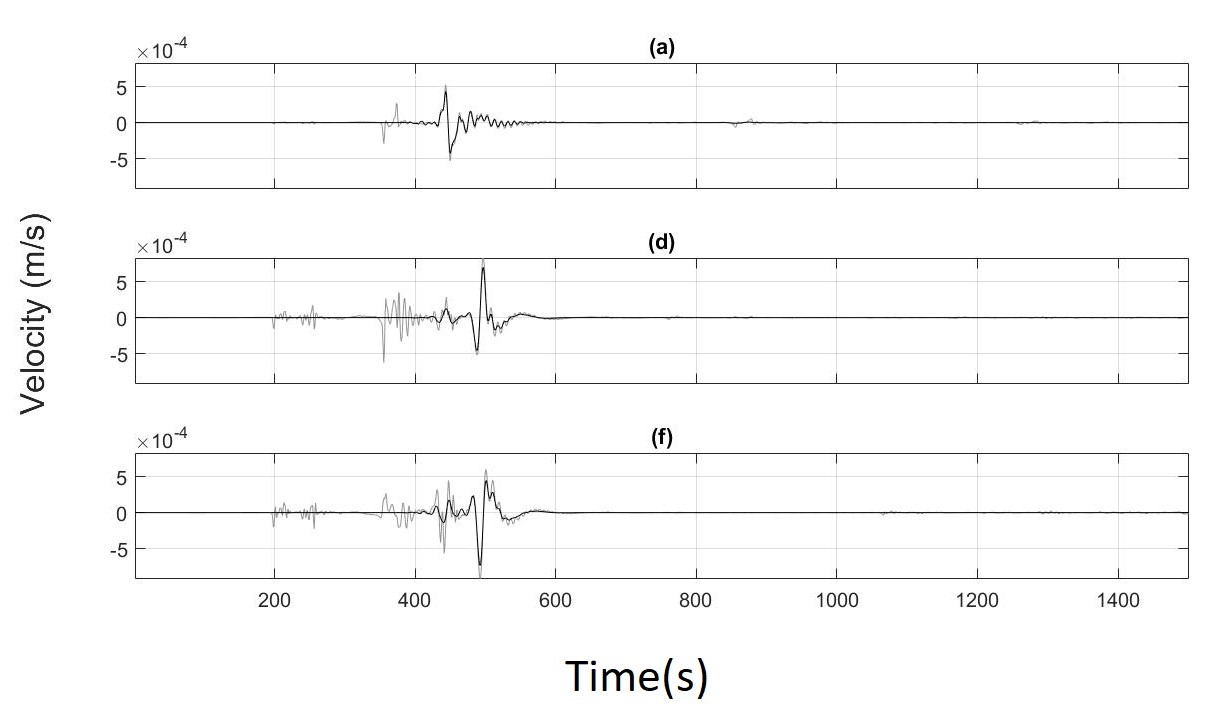} 
		\caption{Body waves separated from surface waves. Panel (a) shows the black waveform of the Love wave, separated using the SP-TFF method applied to the synthetic data. Panels (d) and (f) show the black waveforms of the radial and vertical components of the Rayleigh wave, which have been successfully separated. The gray waveforms in all panels represent the raw synthetic data recorded at station VBMS for the synthetic earthquake, as shown in Figure 2-4.}
		\label{fig:Waveforms232}
	\end{figure}
	
	Another significant challenge is the identification and separation of Love and SH waves, both of which are recorded on the transverse component and often overlap. Considering that body waves have a broader frequency range than surface waves, SH waves typically occupy a wider bandwidth than Love waves. This characteristic can be very helpful for distinguishing SH waves from Love waves in the time--frequency representation. To demonstrate the effectiveness of the SP-TFF method in addressing this challenge, we analyze the data recorded at station TZTN. The figure below shows the locations of the earthquake and this station.

	The waveforms recorded on the transverse, radial, and vertical components for the arrival of the synthetic earthquake at station TZTN are shown in panels (a)--(c) of the figure below.
	\begin{figure}[h!]
		\centering
		\includegraphics[width=1.0\textwidth]{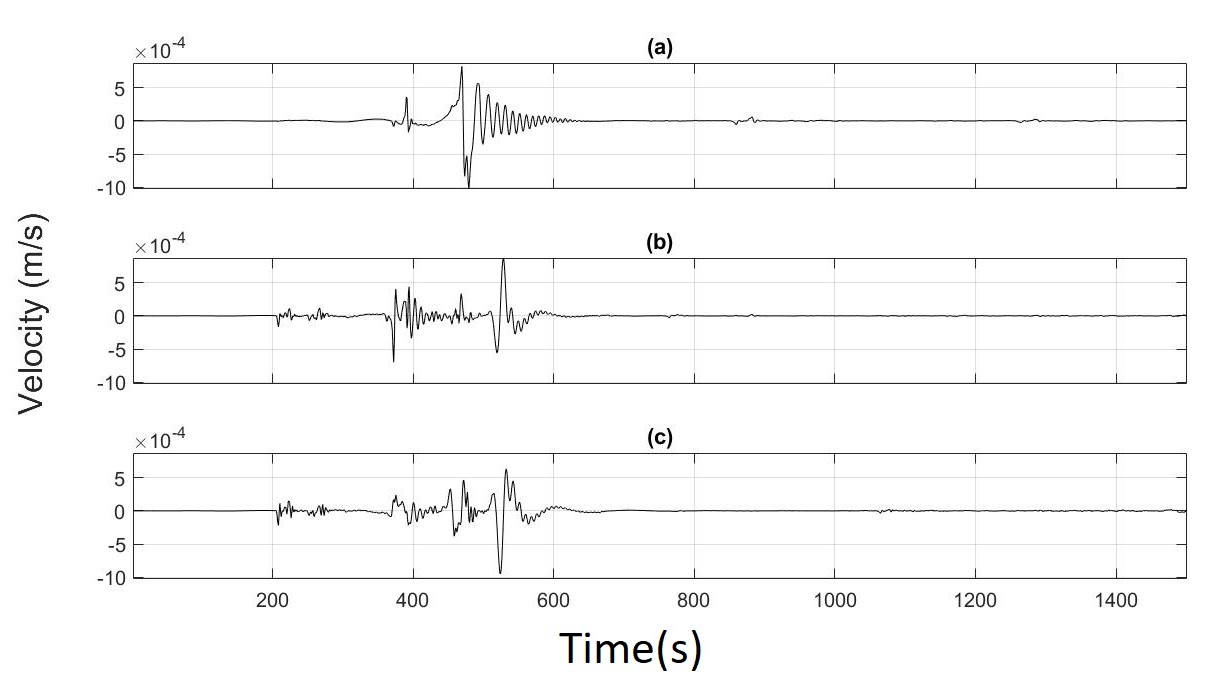} 
		\caption{Waveforms shown in panels (a)--(c) correspond to the transverse, radial, and vertical components of the three-dimensional synthetic earthquake dataset with a moment magnitude of $M_w = 7.0$ in Guerrero, Mexico, generated for station TZTN using the AxiSEM library.}
		\label{fig:Waveforms4}
	\end{figure}
	
	The time--frequency representations of the radial, transverse, and vertical components of the synthetic earthquake data at station TZTN obtained using the ST method are shown in panels (a), (c), and (e) of the figure below. The corresponding time--frequency representations obtained using the SP-TFF method are shown in panels (b), (d), and (f).
	
	\begin{figure}[h!]
		\centering
		\includegraphics[width=1.0\textwidth]{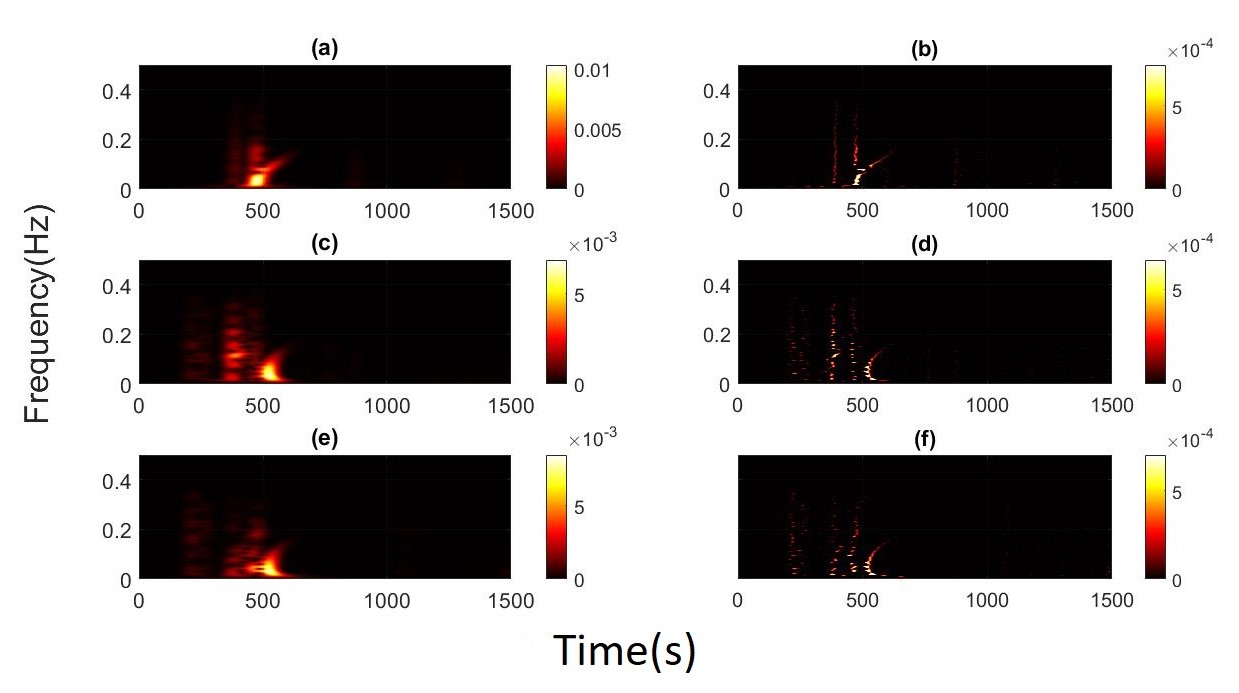} 
		\caption{Time--frequency representations of the synthetic dataset shown in Figure 9-4. Panels (a), (c), and (e) correspond to the radial, transverse, and vertical components obtained using the ST method, while panels (b), (d), and (f) show the corresponding components obtained using the SP-TFF method.}
		\label{fig:Waveforms5}
	\end{figure}  

	In Figure 10-4, panel (b), the SH wave can be observed around 480 seconds, overlapping with the Love wave; we will address the separation of these two waves in the following section.
	
	The figure below shows the time--frequency representations of the major and minor axes of the particle-motion ellipses. Panels (a) and (c) correspond to the $SM$ and $Sm$ axes obtained using Pingar’s method (2006), while panels (b) and (d) show the $SM$ and $Sm$ axes derived from applying EDPA to the SP-TFF method.
	\begin{figure}[h!]
		\centering
		\includegraphics[width=1.0\textwidth]{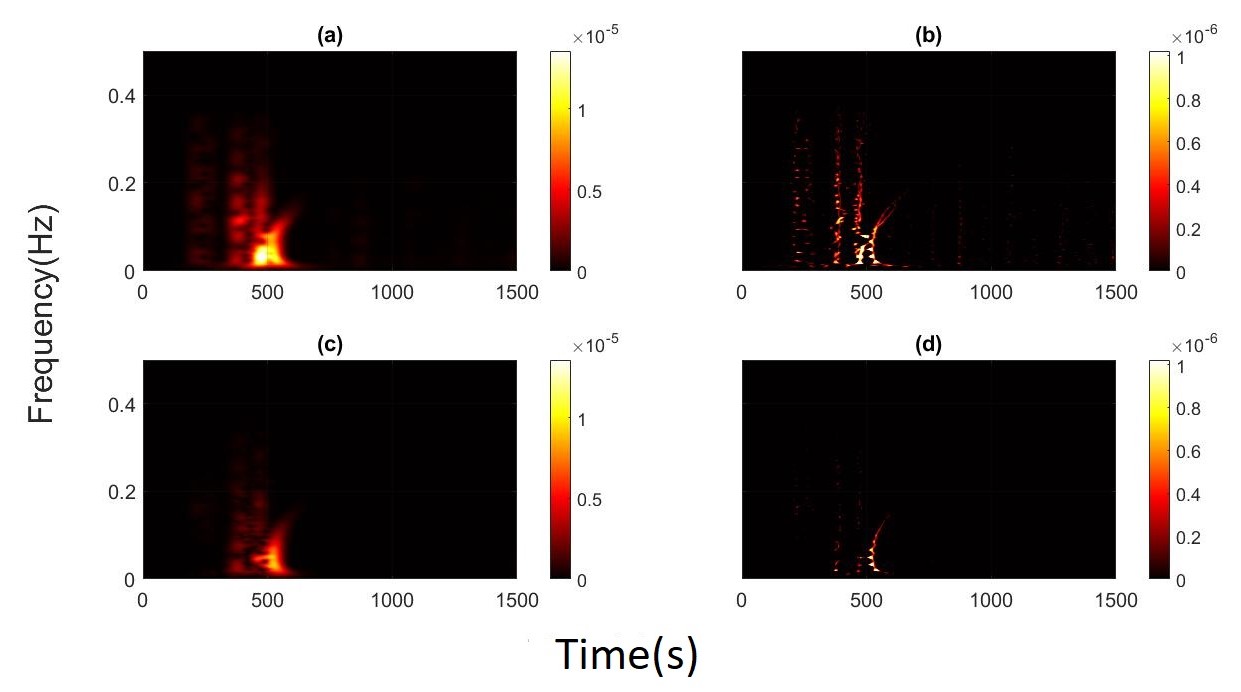} 
		\caption{Time--frequency representations of the major ($SM$) and minor ($Sm$) axes of particle motion for the synthetic dataset shown in Figure 9-4. Panels (a) and (c) correspond to the $SM$ and $Sm$ axes obtained using Pingar’s method (2006), while panels (b) and (d) show the $SM$ and $Sm$ axes derived from applying EDPA to the SP-TFR.}
		\label{fig:Waveforms6}
	\end{figure}

	In this section, by combining the time--frequency domain polarization parameters obtained, the data are processed to filter out the Love-wave phases. For Love-wave filtering, a directivity filter is designed, which defines the directivity measure relative to the transverse axis based on a combination of adaptive parameters. An amplitude filter is also constructed to remove the Love waves according to Equation (3-32). By visual inspection, the approximate arrival time of the Love wave can be determined. This procedure is performed to limit the filtering region and affects both the high-amplitude SH waves and the Love waves.  
	
	The results of applying the Love-wave removal filter to the SP-TFF for the transverse, radial, and vertical components are shown in panels (a), (c), and (e) of Figure 12-4. As observed, the energy corresponding to the Love wave in the time--frequency plane is significantly removed, leaving only the scattered energy related to the body waves. The SP-TFFs of the radial (see Figure 12-4(c)) and vertical (see Figure 12-4(e)) components remain unaffected by the filter.
	
	\begin{figure}[h!]
		\centering
		\includegraphics[width=1.0\textwidth]{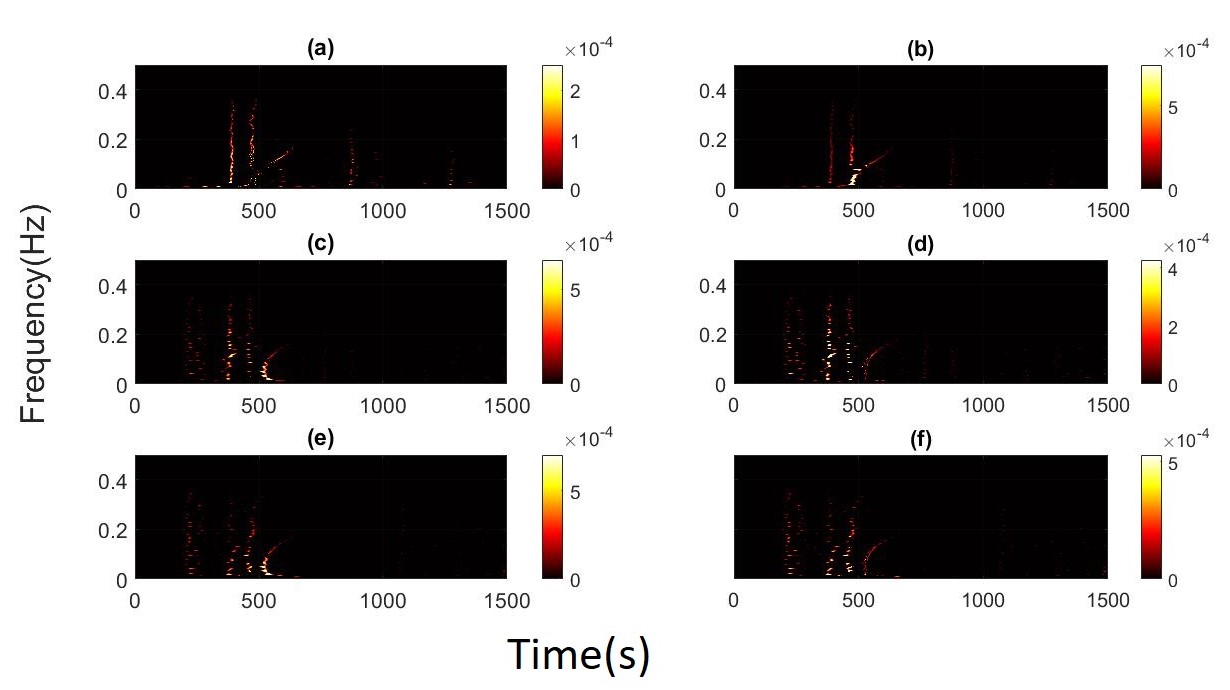} 
		\caption{Adaptive filtering for surface-wave removal and body-wave separation. Panels (a), (c), and (e) show the SP-TFF of Love-wave-filtered data on the transverse, radial, and vertical components, respectively. Panels (b), (d), and (f) display the SP-TFR of Rayleigh-wave-filtered data on the same components.}
		\label{fig:Waveforms7}
	\end{figure} 
	
	The black waveforms in panels (a), (c), and (e) of Figure 13-4 represent the transverse, radial, and vertical components reconstructed in the time domain after filtering. The Love wave is almost completely removed from the transverse component in the time domain, while the SH waves remain in the seismograms. It is observed that the Love-wave filtering process has no effect on other phases in the radial and vertical components. A promising result of time--frequency filtering using the SP-TFF method in this example is that, as shown in panel (a) of the figure below, an SH phase around 480 seconds, which was previously masked by high-amplitude Love waves, is successfully recovered after filtering.
	
	\begin{figure}[h!]
		\centering
		\includegraphics[width=1.0\textwidth]{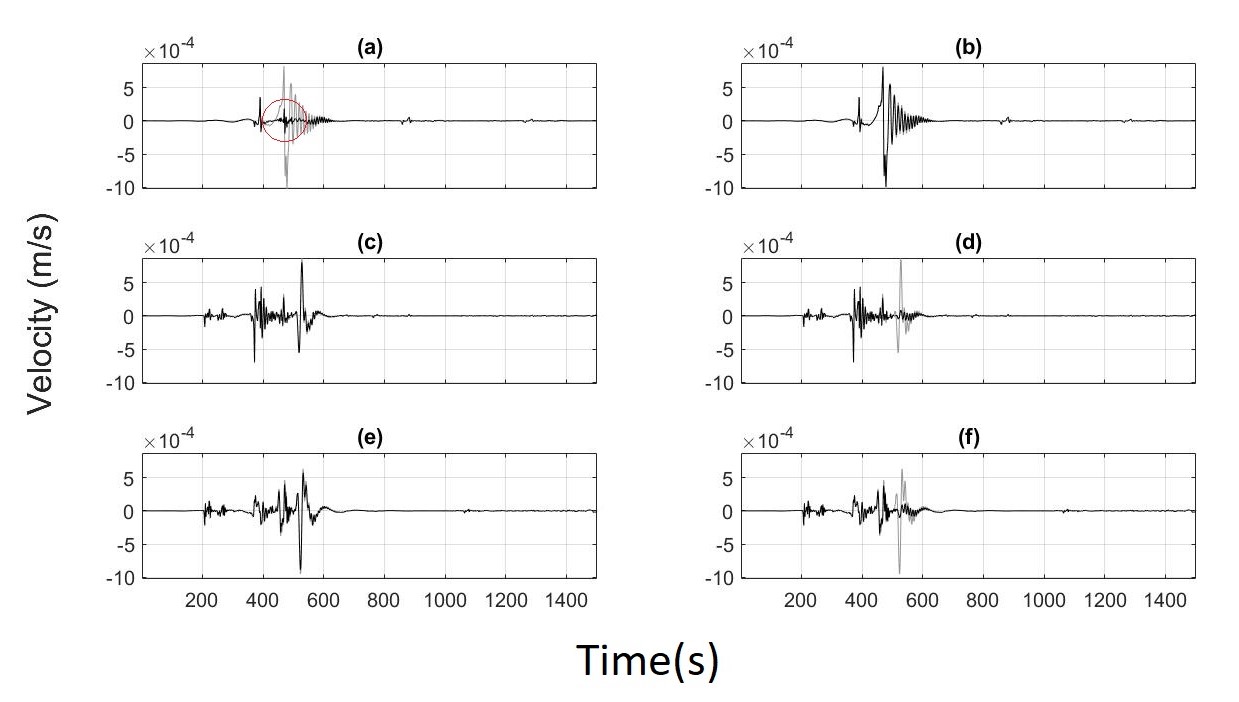} 
		\caption{Filtered surface waves and remaining body waves for the synthetic earthquake. The black waveforms in panels (a), (c), and (e) correspond to the transverse, radial, and vertical components of the Love-wave-filtered synthetic dataset shown in Figure 9-4, obtained using the SP-TFF method. Panels (b), (d), and (f) show the corresponding components for the Rayleigh-wave-filtered data. The gray waveforms in all panels represent the raw synthetic dataset.}
		\label{fig:Waveforms8}
	\end{figure}
	
	As a final assessment, by applying Equation (3-33) to the SP-TFR, the Love-wave phases are extracted on all three components. In Figure 14-4(a), the separated Love wave is clearly visible. Notably, the Love-wave phase has been extracted from the complete seismograms with high accuracy.
	
	\begin{figure}[h!]
		\centering
		\includegraphics[width=1.0\textwidth]{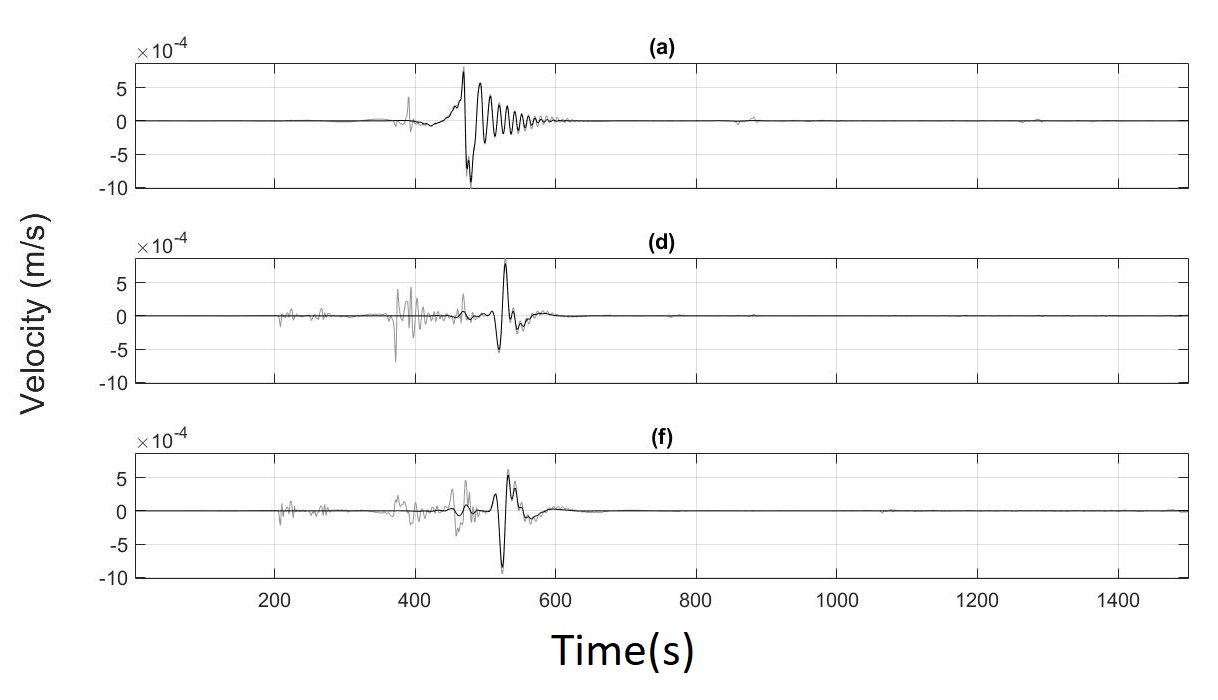} 
		\caption{Body waves separated from surface waves. Panel (a) shows the black waveform of the Love wave, separated using the SP-TFF method applied to the synthetic data. Panels (b) and (c) show the black waveforms of the radial and vertical components of the Rayleigh wave, which have been successfully separated. The gray waveforms in all panels represent the raw synthetic data recorded at station TZNT for the synthetic earthquake, as shown in Figure 9-4.}
		\label{fig:Waveforms9}
	\end{figure}
	
	As shown, the SP-TFF method effectively filters the Love wave without affecting the body waves on the radial and vertical components, and has only a minor effect on other phases on the transverse component.

	\subsection*{Real data examples}
	
	To validate the SP-TFF method, we use the real seismograms of the Guerrero 2021 earthquake  recorded at OGNE station. The waveforms were first processed by applying detrending, time-domain resampling at 2-second intervals, deconvolution of the instrument response, and conversion to velocity. The figure below shows the locations of the earthquake and station OGNE, which is analyzed in this study.	The processed seismograms of station OGNE are shown in Figure 16-4. These seismograms are represented in the transverse--radial--vertical coordinate system. 	The time-frequency representations of these components using both Pingar’s method and SP-TFR are shown in the figure below. Panels (a), (c), and (e) correspond to the transverse, radial, and vertical components obtained using Pingar’s method (2006), while panels (b), (d), and (f) show the time--frequency representations of the same components obtained using the SP-TFR method. As observed for the real data, the time--frequency representation obtained using SP-TFR is significantly more accurate than that obtained using Pingar’s method (2006).
	\begin{figure}[h!]
		\centering
		\includegraphics[width=1.0\textwidth]{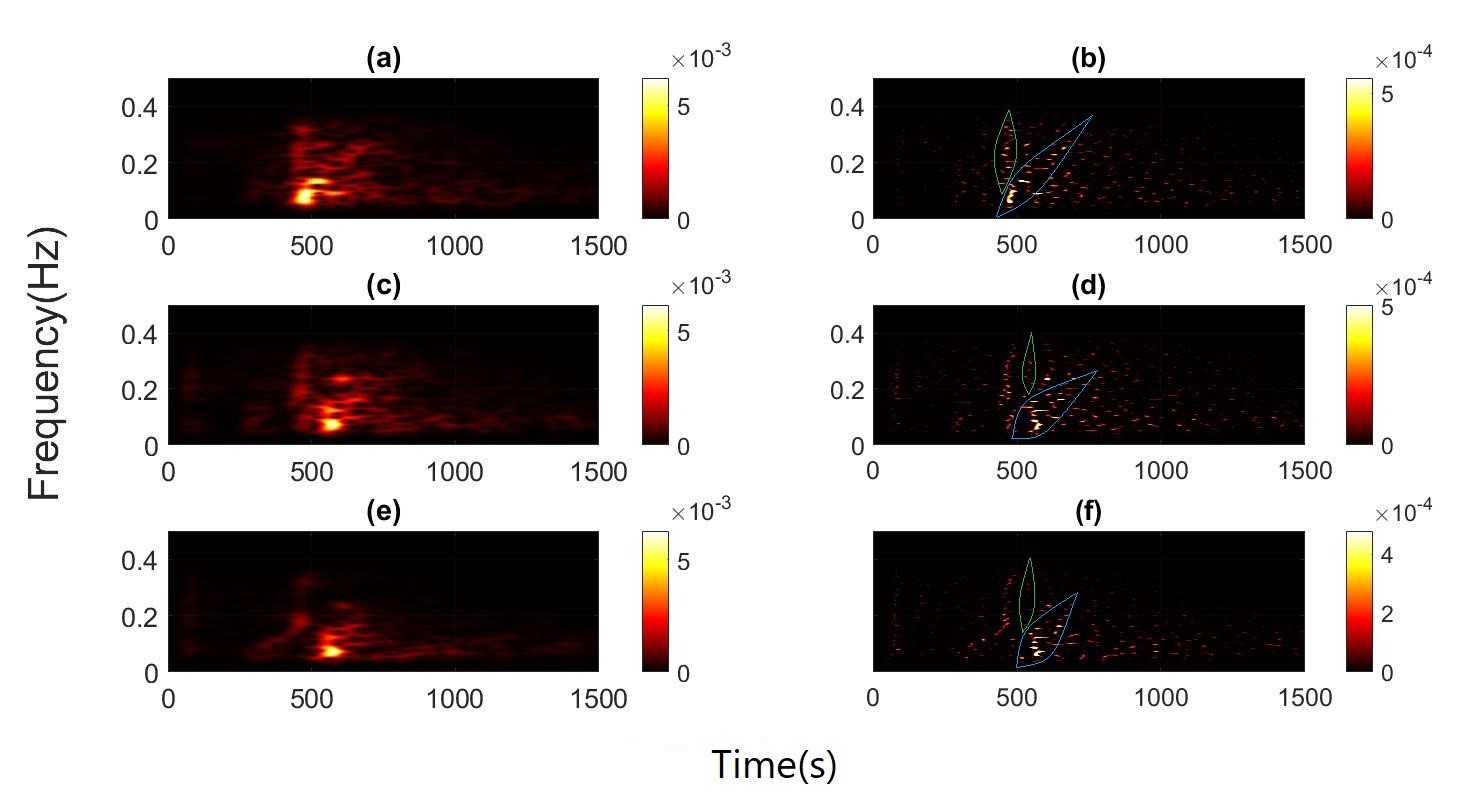} 
		\caption{Time--frequency representations of the real data shown in Figure 16-4. Panels (a), (c), and (e) display the transverse, radial, and vertical components obtained using the ST method, while panels (b), (d), and (f) show the corresponding time--frequency components obtained using the SP-TFR. The regions separated by the green and blue curves represent the body waves and surface waves, respectively, in the time--frequency representation.}
		\label{fig:Waveforms13}
	\end{figure}
	The results are similar to those observed in the synthetic dataset, where SP-TFR (right column) provides a more accurate time--frequency representation compared to the ST method implemented by Pingar (2006) (left column). Although the surface and body waves in the real data do not exhibit a clear pattern as in the synthetic data, they can still be partially distinguished. The high-amplitude regions in panels (b), (d), and (f), whose frequency increases over time, correspond to surface waves, while the body waves exhibit a broader frequency band. In each of these figures, examples of surface and body waves have been delineated using blue and green lines, respectively. To better visualize the separation, we focused on body waves that overlap with surface waves.
	\begin{figure}[h!]
		\centering
		\includegraphics[width=1.0\textwidth]{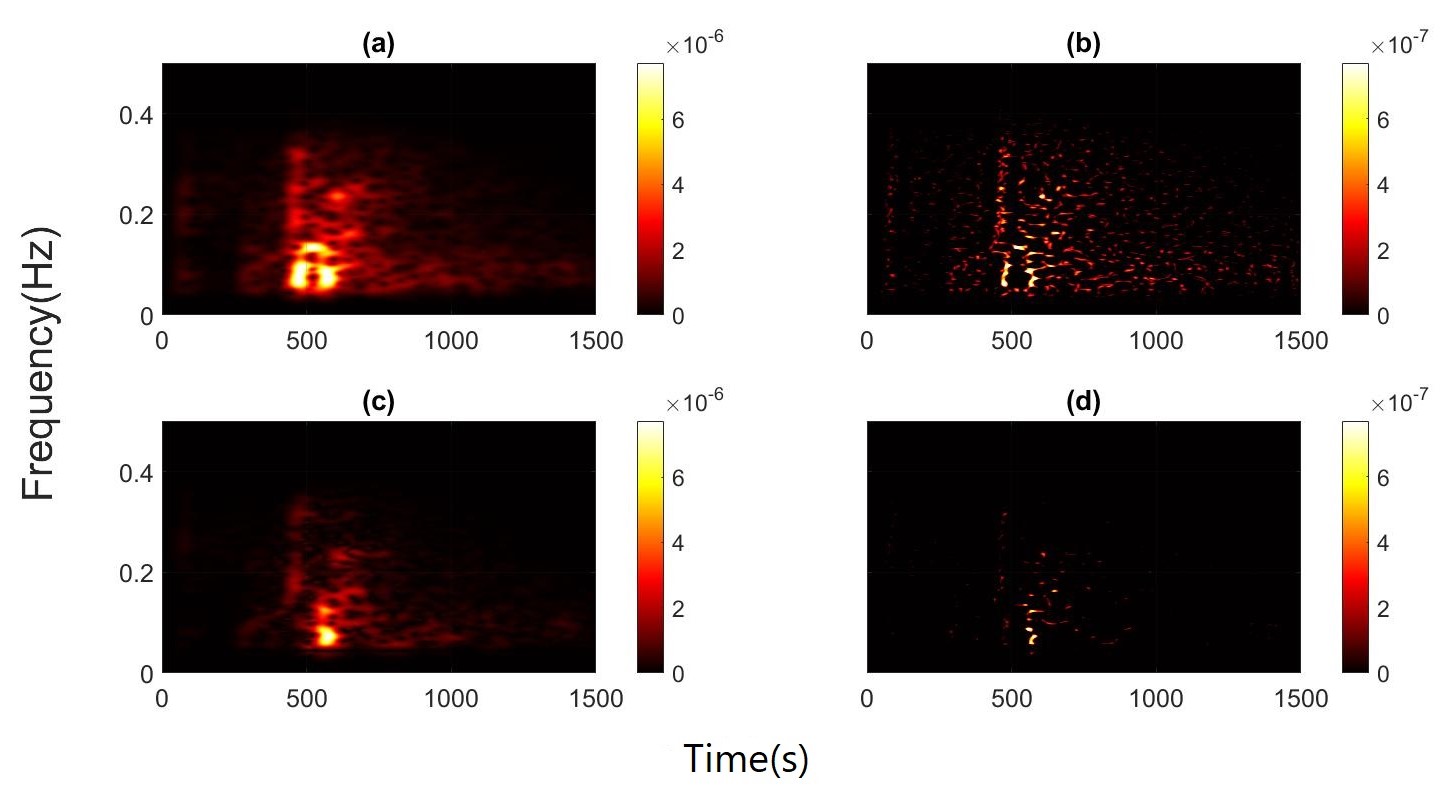} 
		\caption{Time--frequency representations of the $SM$ and $Sm$ axes of particle motion for the real dataset shown in Figure 16-4. Panels (a) and (c) correspond to the $SM$ and $Sm$ axes obtained using Pingar’s method (2006), while panels (b) and (d) show the $SM$ and $Sm$ axes derived from applying EDPA to the SP-TFR. Regions with low amplitude indicate very small $Sm$ values, corresponding to the linear motion of Love and body waves, whereas regions with high amplitude primarily include the elliptical motion of Rayleigh waves, for which $Sm$ is not small.}
		\label{fig:Waveforms14}
	\end{figure}
	
	The distinct polarization patterns between Love and Rayleigh waves are more clearly observed in the $SM$ and $Sm$ axes in the time--frequency domain shown above. The elliptical particle motion of Rayleigh waves can be distinguished from the linear motion of Love and body waves in the $SM$ and $Sm$ axes based on differences in their amplitudes.
	\textbf{Filtering Body Waves Using SP-TFF}  
	
	The filtering process at this stage is carried out in the same manner as described in the synthetic dataset analysis section. The results are presented below.  
	
	The results of applying the Love-wave removal filter to the SP-TFR for the transverse, radial, and vertical components are shown in panels (a), (c), and (e) of Figure 19-4. The SP-TFRs of the radial (see Figure 19-4(c)) and vertical (see Figure 19-4(e)) components remain unaffected by the filter.
	\begin{figure}[h!]
		\centering
		\includegraphics[width=1.0\textwidth]{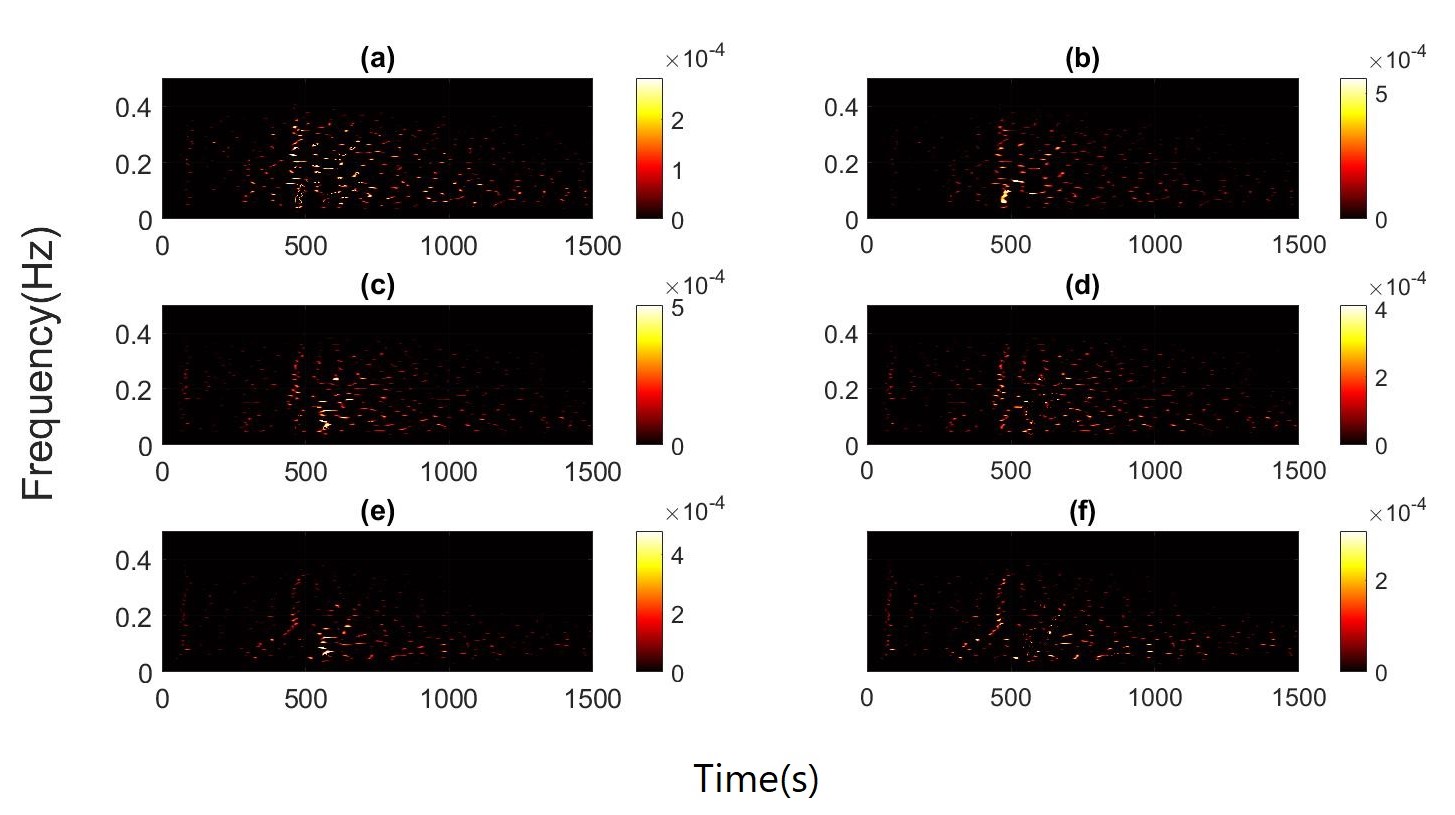} 
		\caption{Adaptive filtering for surface-wave removal and body-wave separation of the real dataset. Panels (a), (c), and (e) show the SP-TFR of Love-wave-filtered data on the transverse, radial, and vertical components, respectively. Panels (b), (d), and (f) display the SP-TFR of Rayleigh-wave-filtered data on the same components.}
		\label{fig:Waveforms15}
	\end{figure}
	
	The transverse component in the figure above corresponds to the separation of body waves from the Love wave. As can be seen by comparing panels (a) and (b), the high-amplitude Love waves around 500 seconds visible in (b) are effectively removed in (a), leaving only the scattered energy of the body waves. To observe the effect of the filter on Rayleigh-wave removal, the radial and vertical components are also examined. Panels (d) and (f) show that, compared to (c) and (e), the Rayleigh waves are well removed while the body waves remain.  
	
	The black waveforms in panels (a), (c), and (e) of the figure below display the transverse, radial, and vertical components of the Love-wave-filtered data in the time domain. These results confirm that the method used in this study specifically removes the Love wave from the transverse component in the time domain without significantly affecting other phases.
	\begin{figure}[h!]
		\centering
		\includegraphics[width=1.0\textwidth]{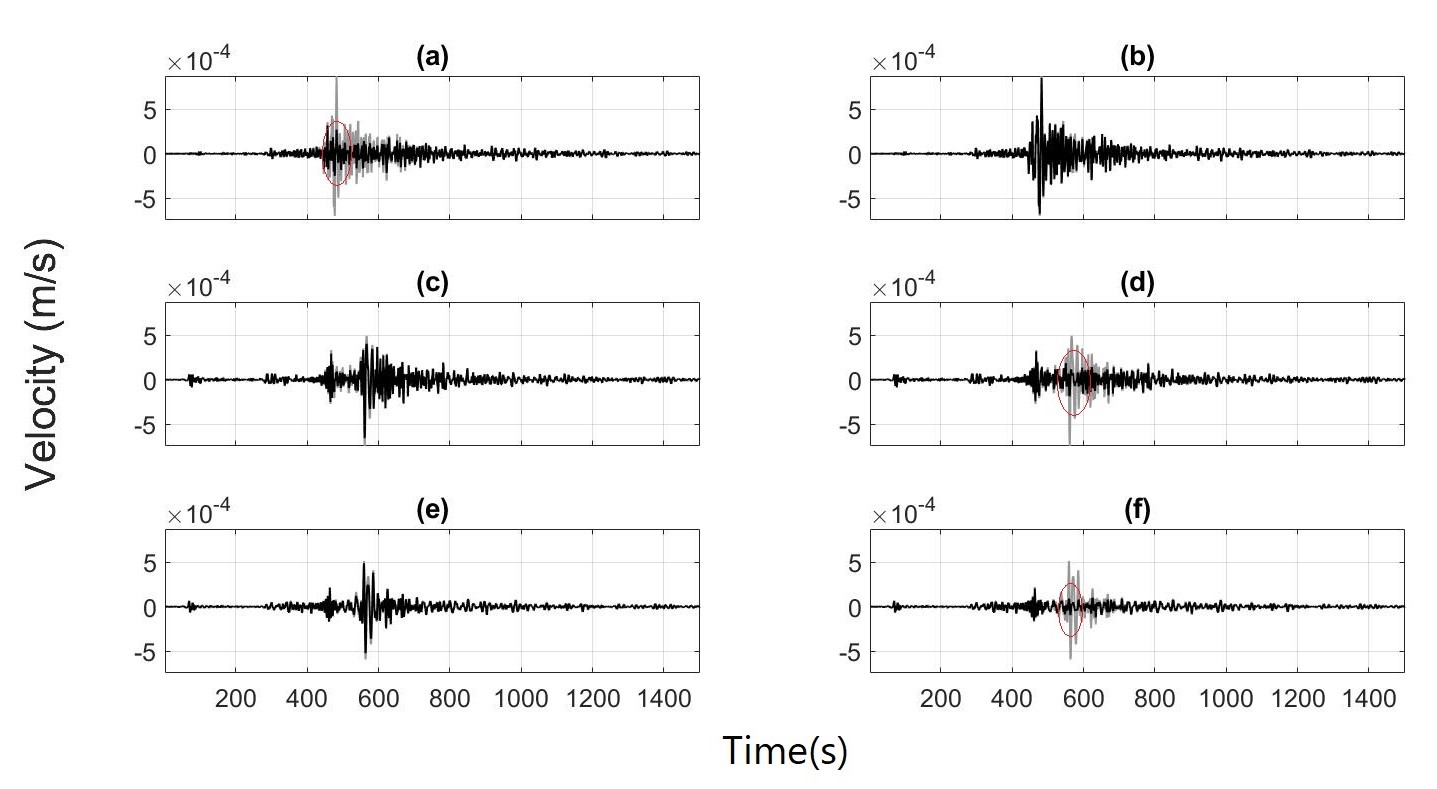} 
		\caption{Surface-wave filtering and remaining body waves of the earthquake. The black waveforms in panels (a), (c), and (e) represent the transverse, radial, and vertical components of the Love-wave-filtered real dataset shown in Figure 16-4, obtained using the SP-TFF method. Panels (b), (d), and (f) show the corresponding Rayleigh-wave-filtered components. The gray waveforms in all panels represent the raw real earthquake data.}
		\label{fig:Waveforms166}
	\end{figure}
	
	Finally, the body-wave phases of the earthquake for the real dataset are separated from the surface waves by applying Equation (3-33) to the SP-TFR of all three components. The separated Love wave and body waves on the transverse component are shown in panel (a) above. Similarly, Figure 21-4(b) and (c) display the radial and vertical components of the Rayleigh and body waves, which have been successfully separated.
		\begin{figure}[h!]
		\centering
		\includegraphics[width=1.0\textwidth]{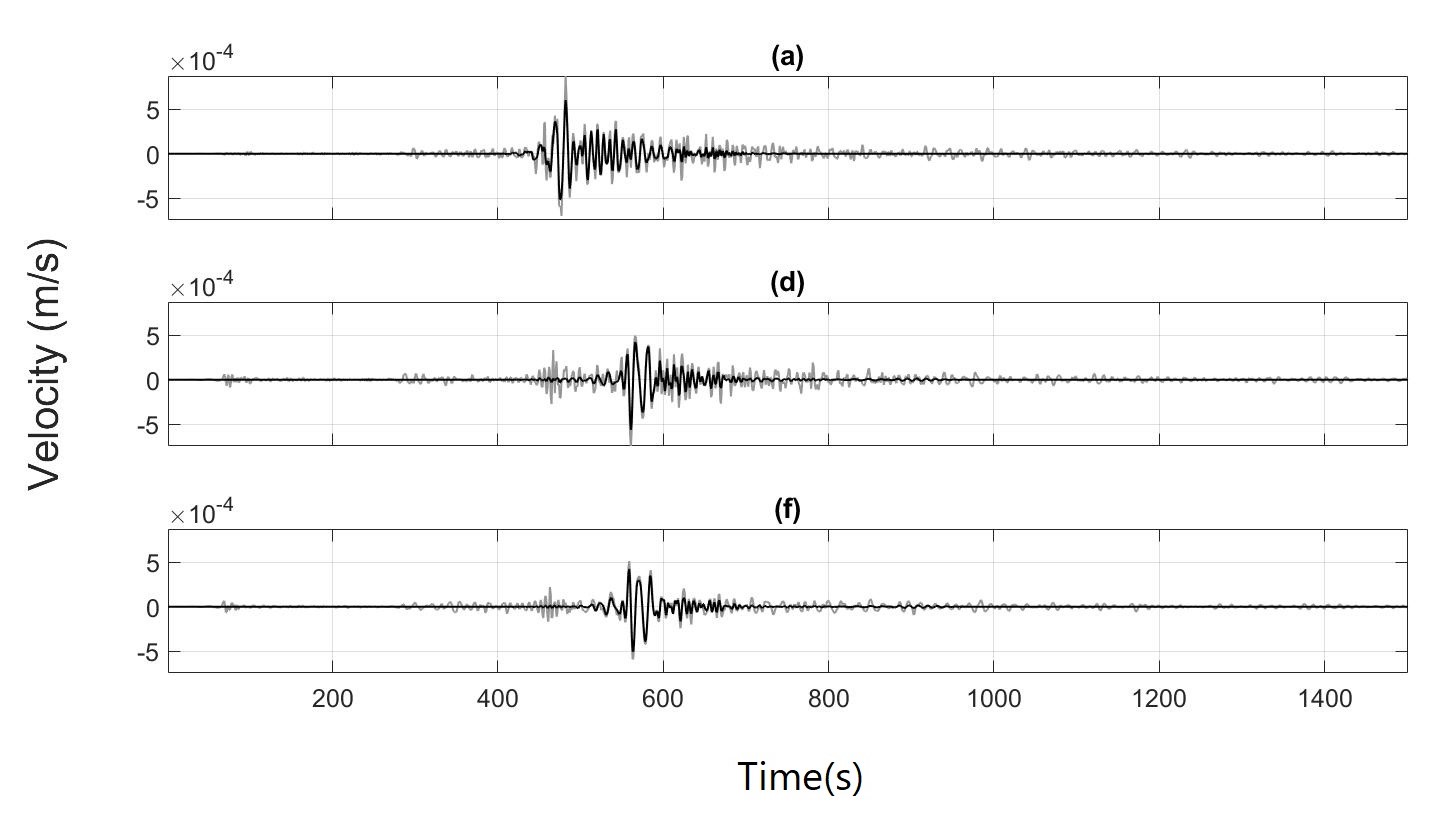} 
		\caption{Body waves separated from surface waves. Panel (a) shows the black waveform of the Love wave, separated from the real data using the SP-TFF method. Panels (b) and (c) display the black waveforms of the radial and vertical components of the Rayleigh wave, which have been successfully separated. The gray waveforms represent the raw data recorded at station OGNE, as shown in Figure 16-4.}
		\label{fig:Waveforms16}
	\end{figure}

\section{Conclusion and future works} 
This study demonstrates that Sparsity-Promoting Time–Frequency Filtering (SP-TFF) provides a powerful framework for the reliable extraction of earthquake body-wave phases that are otherwise obscured by surface-wave energy. By incorporating polarization-based attributes—amplitude, directivity, and rectilinearity—into a tailored filtering strategy, the method achieves focused isolation of body-wave arrivals and suppression of interfering phases. Applications to both synthetic and observational data from the 2021 $M_w 7.0$ Guerrero earthquake confirm that SP-TFF is effective, scalable, and well-suited for automated integration into modern seismological workflows. Beyond improving the resolution of phase extraction, this approach offers a flexible computational tool that can enhance earthquake characterization, support large-scale waveform analyses, and ultimately advance Earth structure imaging. Open access to the codes ensures transparency, reproducibility, and broader adoption within the seismological community. The next step in our study will focus on extracting body waves from Empirical Green’s Functions derived from cross-correlations of ambient seismic noise at local scales.



\FloatBarrier

\begin{appendices}




\end{appendices}

\input{acro_list}

\bibliography{References}

\end{document}

%% file: acro_list.tex
\begin{acronym} 

\acro{WWSSN}{World Wide Standard Seismograph Network}
\acro{GSD}{Geological Survey Department}

\acro{AR}{accuracy rate}
\acro{WAC}{West African Craton}
\acro{MAR}{Mid-Atlantic Ridge}
\acro{SP-TFF}{Sparsity Promoting Time-Frequency Filtering}

\acro{KKT}{Karush–Kuhn–Tucker}

\acro{IRLS}{Iterated Reweighted Least Squares}
\acro{PA}{polarization analysis}
\acro{HR}{high-resolution}

\acro{ROSES}{Remote Online Sessions for Emerging Seismologists}

\acro{TFR}{time-frequency representation}
\acro{SP-TFR}{sparsity-promoting time-frequency representation}
\acro{SP-TFF}{sparsity-promoting time-frequency filtering}
\acro{DOP}{degree of polarization}
\acro{FISTA}{fast iterative soft thresholding algorithm}
\acro{STFT}{short time fourier transform}
\acro{ST}{Stockwell transform}
\acro{WT}{Wavelet transform}
\acro{TSR}{time-scale representation}
\acro{SST}{synchrosqueezing transform}
\acro{SM}{Semi-major}
\acro{Sm}{Semi-minor}
\acro{EDPA}{eigenvalue decomposition polarization analysis}

\acro{SNR}{signal to noise ratio}

\acro{SEM}{Spectral Element Method}

\acro{TF}{Time-Frequency}

\acro{AFZ}{Akwapim Fault Zone}
\acro{CBF}{Coastal Boundary Fault}

\acro{RFZ}{Romanche Fractured Zone}

\end{acronym} 